\documentclass[aps,prd,twocolumn,superscriptaddress,showpacs,preprintnumbers]{revtex4}

\usepackage{graphicx}
\usepackage{amsmath}
\usepackage{amsfonts}
\usepackage{pgf}
\usepackage{multirow}

\newcommand{\B}{{B^0}}
\newcommand{\AB}{{\bar{B}^0}}
\newcommand{\psp}{{\psi'}}
\newcommand{\km}{{K^-}}
\newcommand{\kp}{{K^+}}
\newcommand{\pip}{{\pi^+}}
\newcommand{\pim}{{\pi^-}}
\newcommand{\z}{{Z^-}}
\newcommand{\zm}[1]{{Z(#1)^-}}
\newcommand{\zp}[1]{{Z(#1)^+}}
\newcommand{\zpm}[1]{{Z(#1)^\pm}}
\newcommand{\kst}{{K^*}}
\newcommand{\ee}{{e^+e^-}}
\newcommand{\lp}{{\ell^+}}
\newcommand{\lm}{{\ell^-}}
\newcommand{\lpair}{{\lp\lm}}
\newcommand{\mumu}{{\mu^+\mu^-}}
\newcommand{\psill}{\psp\to\lpair}
\newcommand{\decay}{\B \to \psp \kp \pim}
\newcommand{\cdecay}{\AB \to \psp \km \pip}
\newcommand{\decaykstar}{\B \to \psp (\to \lpair) \kst(\to \kp \pim)}
\newcommand{\decayz}{\B \to \kp \z(\to \psp (\to \lpair) \pim)}
\newcommand{\fb}{\mathrm{fb}^{-1}}
\newcommand{\cm}{\mathrm{cm}}
\newcommand{\mev}{\mathrm{MeV}}
\newcommand{\mevcc}{\mathrm{MeV}/c^2}
\newcommand{\gev}{\mathrm{GeV}}

\newcommand{\gevccsq}{\mathrm{GeV}^2/c^4}
\newcommand{\mbc}{M_{\mathrm{bc}}}
\newcommand{\sx}{M^2_{\km \pip}}
\newcommand{\sy}{M^2_{\psi' \pip}}
\newcommand{\DE}{\Delta E}
\newcommand{\br}{\mathcal{B}}
\newcommand{\Dfun}[6]{D^{#1}_{#2\,#3}(#4, #5, #6)}
\newcommand{\dfun}[4]{d^{#1}_{#2\,#3}(#4)}

\begin{document}




\title{Experimental constraints on the spin and parity of the $\zp{4430}$}

\noaffiliation
\affiliation{University of the Basque Country UPV/EHU, 48080 Bilbao}
\affiliation{University of Bonn, 53115 Bonn}
\affiliation{Budker Institute of Nuclear Physics SB RAS and Novosibirsk State University, Novosibirsk 630090}
\affiliation{Faculty of Mathematics and Physics, Charles University, 121 16 Prague}
\affiliation{University of Cincinnati, Cincinnati, Ohio 45221}
\affiliation{Deutsches Elektronen--Synchrotron, 22607 Hamburg}
\affiliation{Justus-Liebig-Universit\"at Gie\ss{}en, 35392 Gie\ss{}en}
\affiliation{Gifu University, Gifu 501-1193}
\affiliation{II. Physikalisches Institut, Georg-August-Universit\"at G\"ottingen, 37073 G\"ottingen}
\affiliation{Hanyang University, Seoul 133-791}
\affiliation{University of Hawaii, Honolulu, Hawaii 96822}
\affiliation{High Energy Accelerator Research Organization (KEK), Tsukuba 305-0801}
\affiliation{Ikerbasque, 48011 Bilbao}
\affiliation{Indian Institute of Technology Guwahati, Assam 781039}
\affiliation{Indian Institute of Technology Madras, Chennai 600036}
\affiliation{Institute of High Energy Physics, Chinese Academy of Sciences, Beijing 100049}
\affiliation{Institute of High Energy Physics, Vienna 1050}
\affiliation{Institute for High Energy Physics, Protvino 142281}
\affiliation{INFN - Sezione di Torino, 10125 Torino}
\affiliation{Institute for Theoretical and Experimental Physics, Moscow 117218}
\affiliation{J. Stefan Institute, 1000 Ljubljana}
\affiliation{Kanagawa University, Yokohama 221-8686}
\affiliation{Institut f\"ur Experimentelle Kernphysik, Karlsruher Institut f\"ur Technologie, 76131 Karlsruhe}
\affiliation{Korea Institute of Science and Technology Information, Daejeon 305-806}
\affiliation{Korea University, Seoul 136-713}
\affiliation{Kyungpook National University, Daegu 702-701}
\affiliation{\'Ecole Polytechnique F\'ed\'erale de Lausanne (EPFL), Lausanne 1015}
\affiliation{Faculty of Mathematics and Physics, University of Ljubljana, 1000 Ljubljana}
\affiliation{Luther College, Decorah, Iowa 52101}
\affiliation{University of Maribor, 2000 Maribor}
\affiliation{Max-Planck-Institut f\"ur Physik, 80805 M\"unchen}
\affiliation{School of Physics, University of Melbourne, Victoria 3010}
\affiliation{Moscow Physical Engineering Institute, Moscow 115409}
\affiliation{Graduate School of Science, Nagoya University, Nagoya 464-8602}
\affiliation{Kobayashi-Maskawa Institute, Nagoya University, Nagoya 464-8602}
\affiliation{Nara Women's University, Nara 630-8506}
\affiliation{National Central University, Chung-li 32054}
\affiliation{National United University, Miao Li 36003}
\affiliation{Department of Physics, National Taiwan University, Taipei 10617}
\affiliation{H. Niewodniczanski Institute of Nuclear Physics, Krakow 31-342}
\affiliation{Nippon Dental University, Niigata 951-8580}
\affiliation{Niigata University, Niigata 950-2181}
\affiliation{Osaka City University, Osaka 558-8585}
\affiliation{Pacific Northwest National Laboratory, Richland, Washington 99352}
\affiliation{Panjab University, Chandigarh 160014}
\affiliation{University of Pittsburgh, Pittsburgh, Pennsylvania 15260}
\affiliation{Research Center for Electron Photon Science, Tohoku University, Sendai 980-8578}
\affiliation{University of Science and Technology of China, Hefei 230026}
\affiliation{Seoul National University, Seoul 151-742}
\affiliation{Soongsil University, Seoul 156-743}
\affiliation{Sungkyunkwan University, Suwon 440-746}
\affiliation{School of Physics, University of Sydney, NSW 2006}
\affiliation{Tata Institute of Fundamental Research, Mumbai 400005}
\affiliation{Excellence Cluster Universe, Technische Universit\"at M\"unchen, 85748 Garching}
\affiliation{Toho University, Funabashi 274-8510}
\affiliation{Tohoku Gakuin University, Tagajo 985-8537}
\affiliation{Tohoku University, Sendai 980-8578}
\affiliation{Department of Physics, University of Tokyo, Tokyo 113-0033}
\affiliation{Tokyo Institute of Technology, Tokyo 152-8550}
\affiliation{Tokyo Metropolitan University, Tokyo 192-0397}
\affiliation{Tokyo University of Agriculture and Technology, Tokyo 184-8588}
\affiliation{University of Torino, 10124 Torino}
\affiliation{CNP, Virginia Polytechnic Institute and State University, Blacksburg, Virginia 24061}
\affiliation{Wayne State University, Detroit, Michigan 48202}
\affiliation{Yamagata University, Yamagata 990-8560}
\affiliation{Yonsei University, Seoul 120-749}
  \author{K.~Chilikin}\affiliation{Institute for Theoretical and Experimental Physics, Moscow 117218} 
  \author{R.~Mizuk}\affiliation{Institute for Theoretical and Experimental Physics, Moscow 117218}\affiliation{Moscow Physical Engineering Institute, Moscow 115409} 
  \author{I.~Adachi}\affiliation{High Energy Accelerator Research Organization (KEK), Tsukuba 305-0801} 
  \author{H.~Aihara}\affiliation{Department of Physics, University of Tokyo, Tokyo 113-0033} 
  \author{K.~Arinstein}\affiliation{Budker Institute of Nuclear Physics SB RAS and Novosibirsk State University, Novosibirsk 630090} 
  \author{D.~M.~Asner}\affiliation{Pacific Northwest National Laboratory, Richland, Washington 99352} 
  \author{V.~Aulchenko}\affiliation{Budker Institute of Nuclear Physics SB RAS and Novosibirsk State University, Novosibirsk 630090} 
  \author{T.~Aushev}\affiliation{Institute for Theoretical and Experimental Physics, Moscow 117218} 
  \author{T.~Aziz}\affiliation{Tata Institute of Fundamental Research, Mumbai 400005} 
  \author{A.~M.~Bakich}\affiliation{School of Physics, University of Sydney, NSW 2006} 
% ...
\collaboration{The Belle Collaboration}


\begin{abstract}
We perform a full amplitude analysis of $\decay$ decays,
with $\psp\to\mu^+\mu^-\text{ or }e^+e^-$, to constrain the
spin and parity of the $\zm{4430}$. The $J^P=1^+$ hypothesis is
favored over the $0^-$, $1^-$, $2^-$ and $2^+$ hypotheses at the levels
of $3.4\sigma$, $3.7\sigma$, $4.7\sigma$ and $5.1\sigma$, respectively.
The analysis is based on a 711 $\fb$ data sample that contains 
$772\times10^6$ $B\bar{B}$ pairs collected at the $\Upsilon(4S)$ resonance
by the Belle detector at the asymmetric-energy $\ee$ collider KEKB.
\end{abstract}




\maketitle


\section{Introduction}

Recently, a number of new states containing a $c\bar{c}$ quark pair have been
observed, many of which are not well described by
the quark model~\cite{olsen_godfrey, brambilla}.
Among these states are charged charmonium-like state candidates;
their minimal quark content is necessarily exotic: 
$|c\bar{c}u\bar{d}\rangle$.
The Belle Collaboration observed a resonance-like structure, the
$\zp{4430}$, in the $\psp\pip$ invariant mass spectrum in
$\bar{B}^0\to\psp\km\pip$ decays~\cite{choiolsen, z4430dalitz}.
Two resonance-like structures, the $\zp{4050}$ and $\zp{4250}$,
were observed in the $\chi_{c1}\pip$ invariant mass spectrum in
$\bar{B}^0\to\chi_{c1}\km\pip$ decays~\cite{mizukchistov}.
The BaBar collaboration searched for these states in
$\bar{B}^0\to\psp\km\pip$ and $\bar{B}^0\to J/\psi\km\pip$ decays~
\cite{babarjpkpi} and in $\bar{B}^0\to\chi_{c1}\km\pip$
decay~\cite{babarchickpi} but did not confirm them.
The BESIII and Belle Collaborations also observed
the $\zpm{3900}$ in the $J/\psi \pi^\pm$
invariant mass spectrum in $Y(4260)\to J/\psi \pi^+ \pi^-$
decays~\cite{z3900bes, z3900belle}.

The results described in Ref.~\cite{z4430dalitz} are based on
a two-dimensional Dalitz analysis.
Here we present the results of a full amplitude analysis of the same
decay $\decay$, with $\psp\to\mu^+\mu^-\text{ or }e^+e^-$;
the decay channel $\psp\to J/\psi \pi^+ \pi^-$ is omitted
due to higher multiplicity of the final state.
The full amplitude analysis is more sensitive to the $\zm{4430}$
quantum numbers than a Dalitz analysis because there is no information
loss due to integration over angular variables.
The analysis is performed using a $711\ \fb$ data sample
collected by the Belle detector
at the asymmetric-energy $\ee$ collider KEKB~\cite{kekb}.
The data sample was collected at the
$\Upsilon(4S)$ resonance and contains $772\times10^6$ $B\bar{B}$ pairs.


\section{The Belle Detector}

The Belle detector is a large-solid-angle magnetic
spectrometer that consists of a silicon vertex detector (SVD),
a 50-layer central drift chamber (CDC), an array of
aerogel threshold Cherenkov counters (ACC),
a barrel-like arrangement of time-of-flight
scintillation counters (TOF), and an electromagnetic calorimeter
comprised of CsI(Tl) crystals (ECL) located inside 
a superconducting solenoid coil that provides a 1.5~T
magnetic field.  An iron flux return located outside of
the coil is instrumented to detect $K_L^0$ mesons and to identify
muons (KLM).  The detector
is described in detail elsewhere~\cite{Belle}.
Two inner detector configurations were used. A 2.0 cm beampipe
and a 3-layer silicon vertex detector were used for the first sample
of 140 $\fb$, while a 1.5 cm beampipe, a 4-layer
silicon detector and a small-cell inner drift chamber were used to record  
the remaining 571 $\fb$\cite{svd2}.  

We use a GEANT-based Monte Carlo (MC) simulation~\cite{geant} to model
the response of the detector, identify potential backgrounds and
determine the acceptance. The MC simulation includes run-dependent
detector performance variations and background conditions.
Signal MC events are generated with Evtgen~\cite{evtgen}
in proportion to the relative luminosities of the
different running periods.


\section{Event selection}

We select events of the type $\decay$ (inclusion of charge-conjugate modes
being implied), where the $\psp$ meson is reconstructed via its $\ee$ and
$\mumu$ decay channels.

All tracks are required to originate from the interaction point region,
$dr < 0.2\ \cm$ and $|dz| < 2\ \cm$, where $dr$ and $dz$ are the cylindrical
coordinates of the point of closest approach of the track to the beam axis.
The $z$ axis of the reference frame coincides with the positron beam axis;
its origin is the interaction point.
Charged $\pi$ and $K$ mesons are identified using
likelihood ratios $R_{\pi/K} = \mathcal{L}_\pi/(\mathcal{L}_\pi+\mathcal{L}_K)$
and $R_{K/\pi} = \mathcal{L}_K/(\mathcal{L}_\pi+\mathcal{L}_K)$,
where $\mathcal{L}_\pi$ and $\mathcal{L}_K$ are the likelihoods for
$\pi$ and $K$, respectively, that are calculated from the combined
time-of-flight information from the TOF, the number of photoelectrons from
the ACC and $dE/dx$ measurements in the CDC. We require $R_{\pi/K}>0.6$
for $\pi$ candidates and $R_{K/\pi}>0.6$ for $K$ candidates.
The $K$ identification efficiency is typically 90\% and
the misidentification probability is about 10\%. Muons are identified by
their range and transverse scattering in the KLM.
Electrons are identified by the presence of a matching
electromagnetic shower in the ECL. An electron
veto is imposed on $\pi$ and $K$ candidates.

For  $\psp\to\ee$ candidates, we include photons that have energies greater
than 30 $\mev$ and are within 50 mrad of the lepton direction
in the calculation of the $\psp$ invariant mass.
We require $|M(\lpair)-m_{\psp}|<60\ \mevcc$, where $\ell$
is either $\mu$ or $e$.
We perform a mass-constrained fit to the $\psp$ candidates.

The beam-energy-constrained mass of the $B$ meson is defined as
$\mbc = \sqrt{E_{\mathrm{beam}}-(\sum_i\vec{p}_i)^2}$,
where $E_{\mathrm{beam}}$ is the
beam energy in the center-of-mass frame and $\vec{p}_i$ are momenta
of decay products in the same frame. We require
$|\mbc-m_B| < 7\ \mevcc$, where $m_B$ is the $B^0$ mass~\cite{PDG}.
A mass-constrained fit is applied to the $B$ meson candidates.

\section{Event distributions and signal yield}

\begin{figure}[ht]
\begin{center}
\includegraphics[width=6cm]{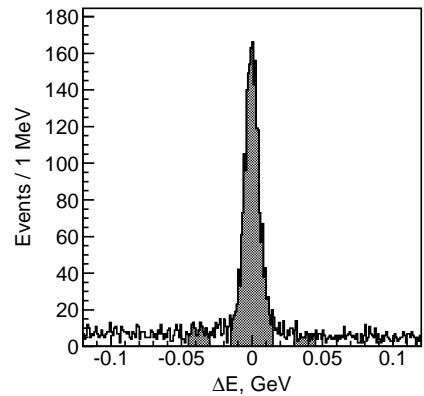}
\end{center}
\caption{The $\DE$ distribution; the signal and sideband regions are hatched.}
\label{fig:deltae}
\end{figure}

The difference between
the reconstructed energy and the
beam energy $\DE = \sum_i E_i - E_{\mathrm{beam}}$,
where $E_i$ are energies of the $B^0$ decay products, is
used to identify the signal.
The signal region is defined as $|\DE| < 15\ \mev$, and the sidebands
are defined as $30\ \mev < |\DE| < 45\ \mev$.
The $\DE$ distribution is shown in
Fig.~\ref{fig:deltae}. 

To determine the signal and background event yields, we perform a binned
maximum likelihood fit of the $\DE$ distribution. It is fitted to
the sum of two Gaussian functions to represent the signal and a second-order
polynomial for the background; all parameters are free. 
The total number of events in the signal region is $2181$;
the number of signal events in the signal region is determined to be
$2010\pm50\pm40$ (here and elsewhere in the paper,
the first uncertainty is statistical and the second is systematic).
Systematic errors are estimated by changing
the $\DE$ fit interval and the order of the polynomial.
We find multiple candidates in 1.4\% of events; no best candidate
selection is applied.

\begin{figure}[ht]
\begin{center}
\includegraphics[width=6cm]{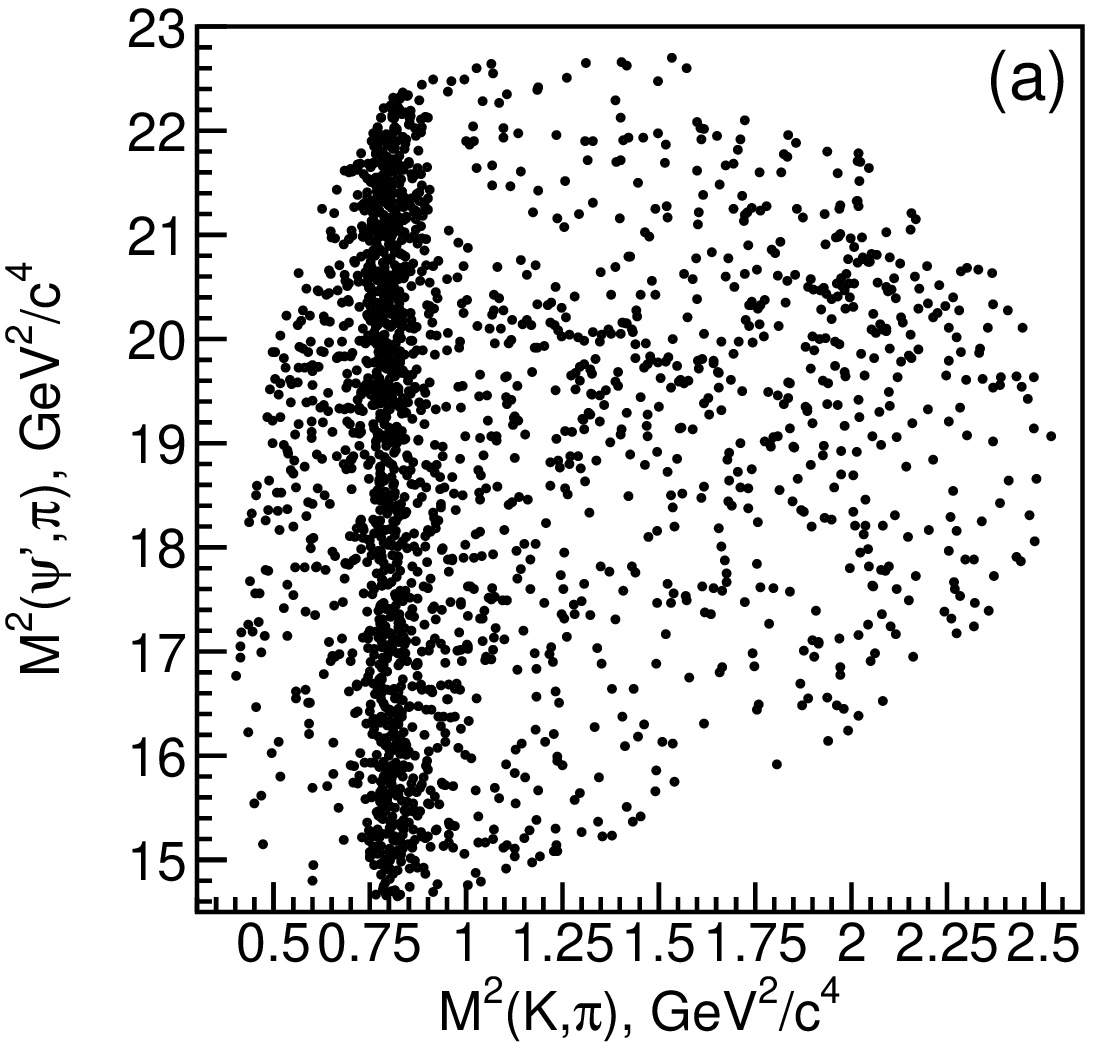}
\includegraphics[width=6cm]{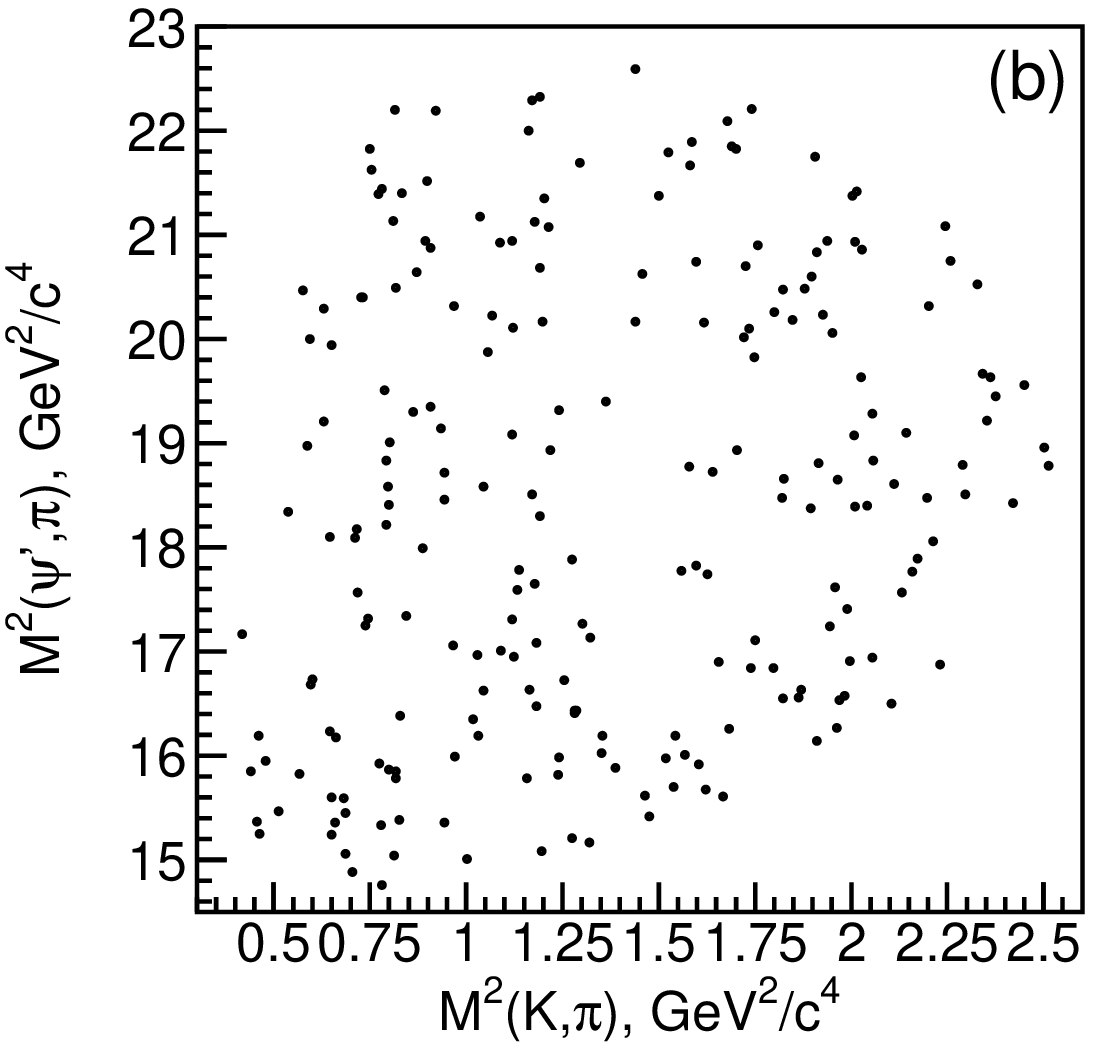}
\includegraphics[width=6cm]{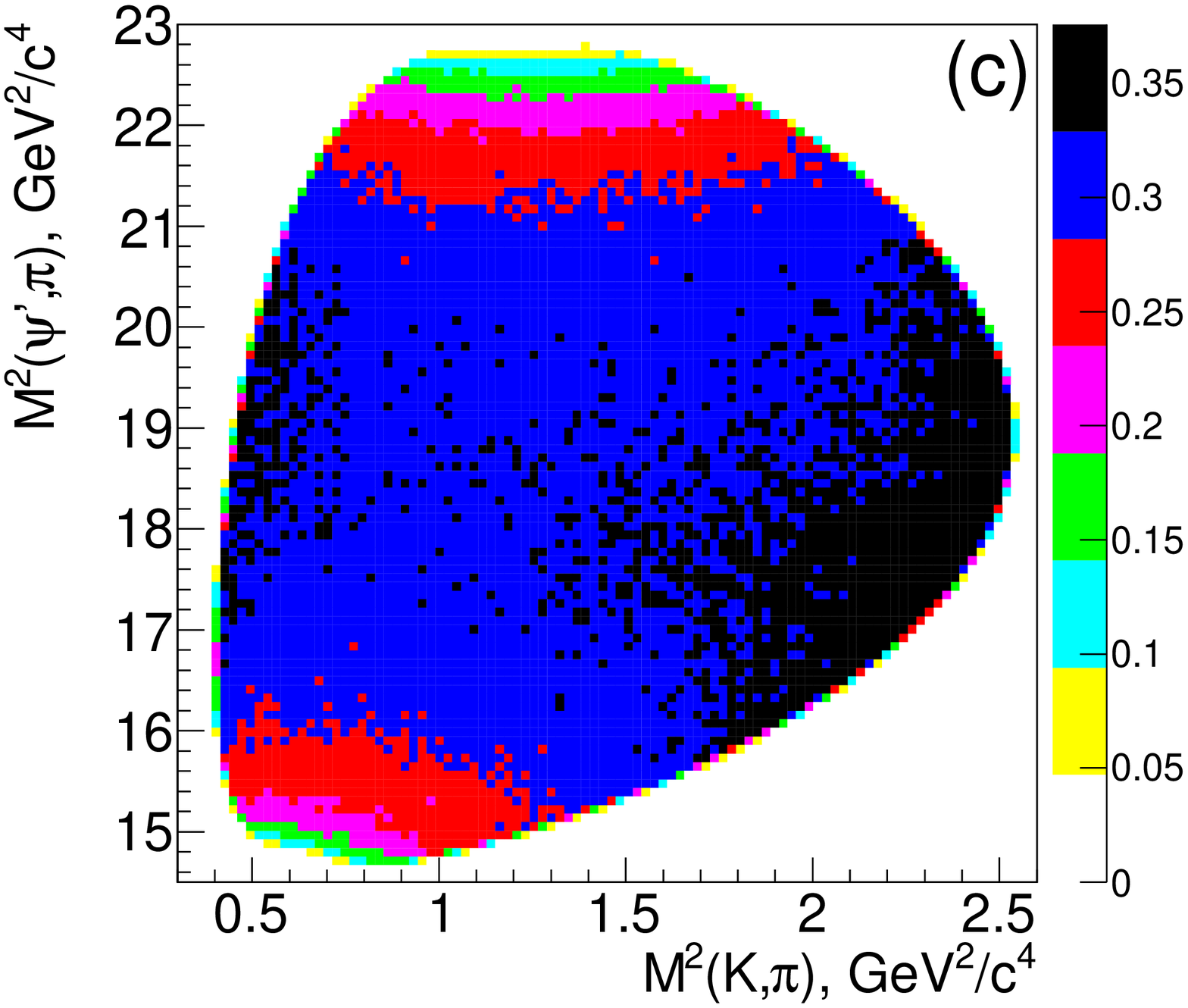}
\end{center}
\caption{The Dalitz plots of the signal region (a), sidebands (b) and
efficiency (c).}
\label{fig:dalitz}
\end{figure}

The Dalitz distribution of $M^2_{\km\pip}$ vs $M^2_{\psp\pip}$ for the signal
region is shown in Fig.~\ref{fig:dalitz}(a).
The vertical band due to production of the intermediate $K^*(892)$
resonance is clearly visible. The Dalitz distribution in
Fig.~\ref{fig:dalitz}(b) for the sidebands is featureless.

To calculate the reconstruction efficiency, we generate MC events for
$\B \to \psp (\to \lp \lm) \kp \pim$ with a uniform
phase space distribution. The efficiency is corrected for the difference
between the particle identification efficiency in
data and MC, which is obtained from a $D^{*+}\to D^0(\to K^-\pi^+)\pi^+$
control sample for $K$ and $\pi$ and a sample of $\gamma\gamma\to\lpair$
for $\mu$ and $e$.

\begin{figure}[ht]
\begin{center}
\includegraphics[width=4.2cm]{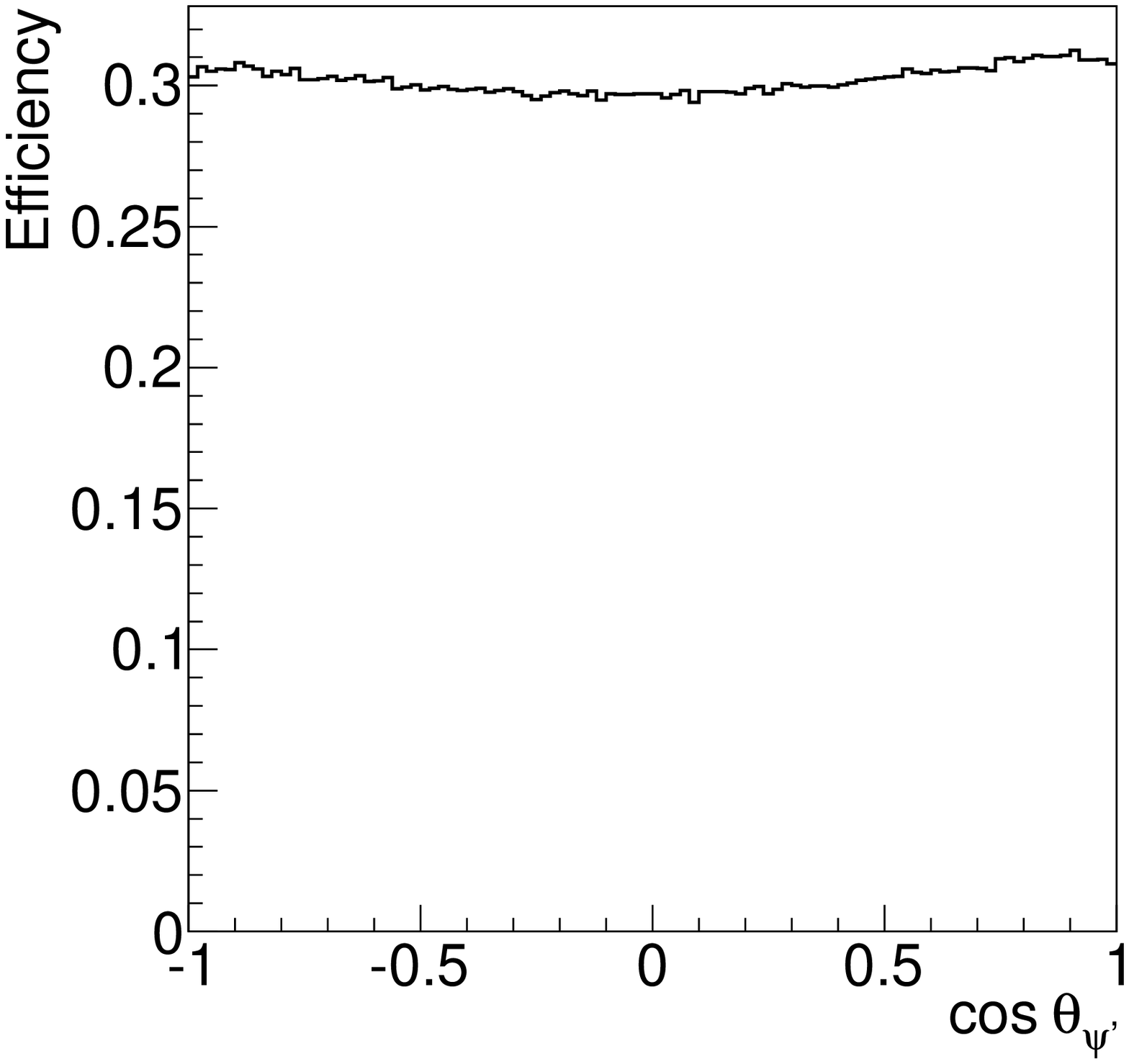}
\includegraphics[width=4.2cm]{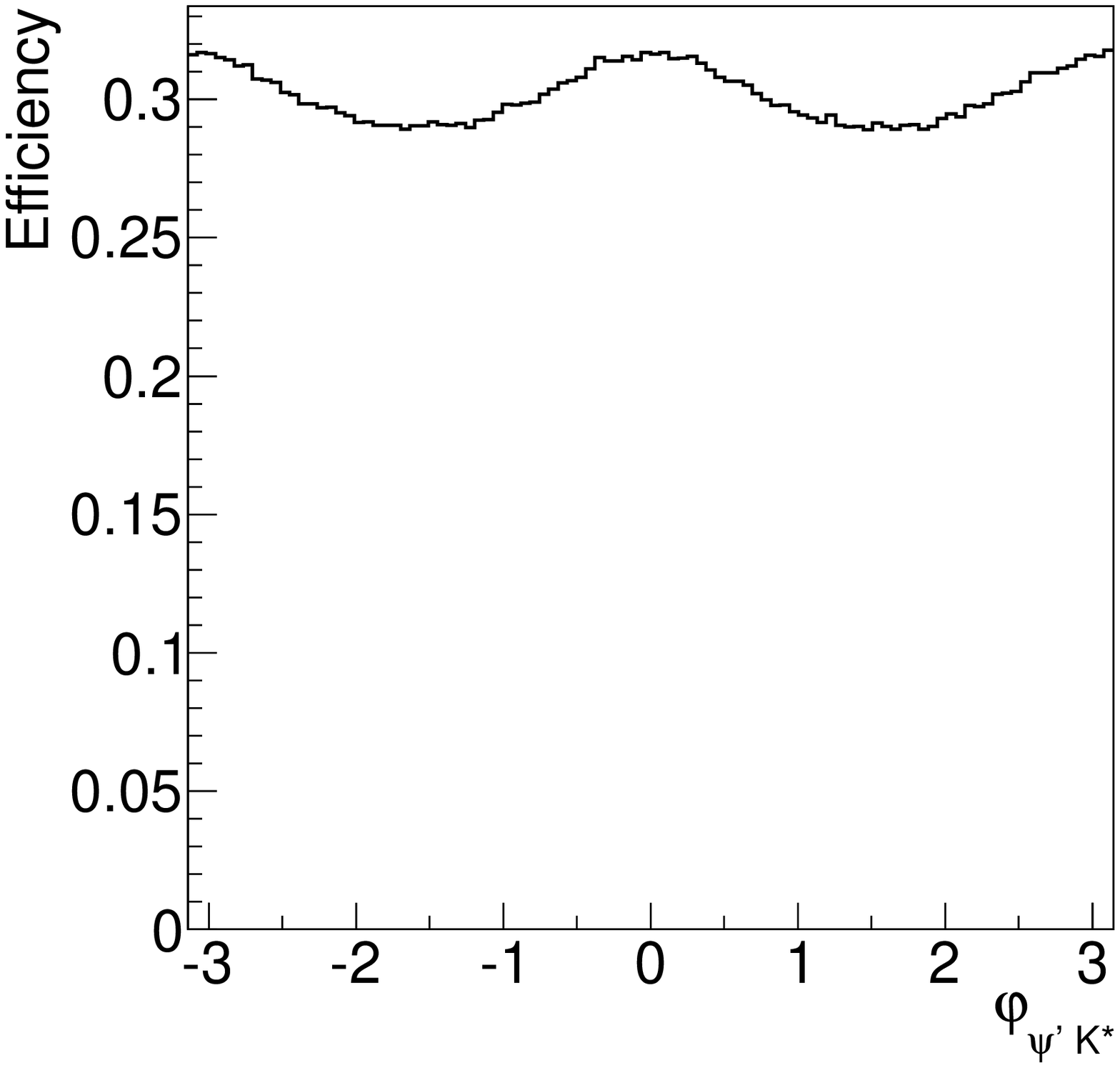}
\end{center}
\caption{Efficiency as a function of angular variables.}
\label{fig:angeff}
\end{figure}

The efficiency as a function of the Dalitz
variables is shown in Fig.~\ref{fig:dalitz}(c). The efficiency drops in the
lower left corner due to slow pions and in the upper corner due to slow kaons;
elsewhere it is almost flat.
The efficiency as a function of the angular variables is shown
in Fig.~\ref{fig:angeff}; $\theta_{\psp}$ is the $\psp$ helicity angle
[the angle between the momenta of the $(\kp,\pim)$ system and the $\mu^-$
in the $\psp$ rest frame] and
$\varphi$ is the angle between the planes defined by
the $(\lp,\lm)$ and $(\kp,\pim)$ momenta
in the $B^0$ rest frame.
The efficiency variation in these distributions is at the 10\% level.

\section{Amplitude analysis formalism}

The amplitude of the decay $\decay$ is
represented by the sum of Breit-Wigner contributions for several
intermediate two-body states.
Our default fit model includes all known $\kp\pim$ resonances below
the kinematic boundary ($1593\ \mevcc$)
[$\kst_0(800)$, $\kst(892)$, $\kst(1410)$, $\kst_0(1430)$,
$\kst_2(1430)$], the first resonance above the boundary [$\kst(1680)$],
and an exotic $\psp\pip$ resonance.

The amplitude is calculated in a four-dimensional parameter space,
defined by
\begin{equation}
\Phi = (\sx,\,\sy,\,\theta_{\psp},\,\varphi).
\end{equation}
The angle-independent part of the amplitude for the decay $\decay$ via a
two-body intermediate resonance $R$ (where $R$ denotes either a $\kp\pim$ or
$\psp\pim$ resonance) is given by
\begin{equation} 
\label{eq:ampl}
\begin{split}
A^R(M^2_R)=
\frac{F^{(L_B)}_B F^{(L_R)}_R
\left(\frac{p_B}{m_B}\right)^{L_B} \left(\frac{p_R}{M_R}\right)^{L_R}}
{m^2_R-M^2_R-i m_R\Gamma(M_R)},
\end{split}
\end{equation}
where $M_R$ is the invariant mass of two daughters of the $R$ resonance;
$F^{(L_B)}_B$ and $F^{(L_R)}_R$ are the $\B$ meson and $R$
resonance decay form factors (the superscript denoting the orbital
angular momentum of the decay);
$(p_B / m_B)^{L_B}\cdot(p_R / M_R)^{L_R}$ is
related to the momentum dependence of the wave function, with $p_B$
($p_R$) being the $\B$ meson ($R$ resonance) daughter's momentum in the
$B$ ($R$) rest frame; $m_B$ is the $\B$ meson mass;
$m_R$ is the mass and $\Gamma(M_R)$ is the energy-dependent
width of the $R$ resonance.
The formula~\eqref{eq:ampl} is the same as in the previous Belle analyses
~\cite{z4430dalitz,mizukchistov}.
The angle-independent part of the nonresonant
amplitude is given by Eq.~\eqref{eq:ampl} without the denominator.

We use the Blatt-Weisskopf form factors~\cite{blattweisskopf}:
\begin{equation}
\begin{split}
F^{(0)}= & 1 ,\\
F^{(1)}= & \sqrt{\frac{1+z_0}{1+z}} ,\\
F^{(2)}= & \sqrt{\frac{z_0^2+3z_0+9}{z^2+3z+9}} ,\\
F^{(3)}= & \sqrt{\frac{z_0^3+6z_0^2+45z_0+225}{z^3+6z^2+45z+225}},\\
\end{split}
\end{equation}
where $z=r^2p_R^2$ ($r$ being the hadron scale)
and $z_0=r^2p_{R0}^2$, where $p_{R0}$ is the $R$
resonance daughter's momentum calculated for the pole mass of the $R$
resonance.
For $\kst$ resonances with nonzero spin $J$, the $B$ decay orbital
angular momentum $L_B$ can have the values $J-1$, $J$ and $J+1$.
 We take the lowest allowed $L_B$ as the default value and
consider the other possibilities in the systematic uncertainty.  The
energy-dependent width is parametrized as
\begin{equation}
\Gamma(M_R)=\Gamma_0\cdot(p_R/p_{R0})^{2L_R+1}\cdot(m_R/M_R)\cdot F_R^2.
\end{equation}

The angle-dependent part of the amplitude is obtained using the helicity
formalism. The amplitude of the decay $\decaykstar$ for one $\kst$ resonance is
\begin{equation}
\begin{split}
A_{\lambda\,\xi}^{\kst}(\Phi) = & H^{\kst}_{\lambda}
A^{\kst}(\sx)\, \dfun{J(\kst)}{\lambda}{0}{\theta_\kst} \\
&\times e^{i\lambda\varphi}\dfun{1}{\lambda}{\xi}{\theta_\psp},
\end{split}
\label{eq:ampkst}
\end{equation}
where $\lambda$ is the helicity of the $\psp$ (the quantization axis being
parallel to the $\kst$ momentum in the $\psp$ rest frame);
$\xi$ is the helicity of the lepton pair;
$H^R_{\lambda}$ is the helicity amplitude for the decay via the intermediate
resonance $R$;
$d^{J(\kst)}_{\lambda\,0}$ and $d^1_{\lambda\,\xi}$ are Wigner $d$ functions;
$J(\kst)$ is the spin of the $\kst$ resonance and $\theta_{\kst}$ is the
$\kst$ helicity angle (the angle between the $\psp$
and $\pim$ momenta in the $\kst$ rest frame).
For $K^*$ resonances with spin 0,
only $\lambda=0$ is allowed. The angle-dependent part of the
nonresonant amplitude is given by Eq.~\eqref{eq:ampkst} with relative angular momentum between the $\kp$ and $\pim$ instead of $J(\kst)$.

For the decay $\decayz$, the amplitude is
\begin{equation}
\begin{split}
A_{\lambda'\,\xi}^{\z}(\Phi) = & H^{\z}_{\lambda'}
A^{\z}(\sy)\,
\dfun{J(\z)}{0}{\lambda'}{\theta_\z} \\
& \times e^{i\lambda'\tilde{\varphi}}
\dfun{1}{\lambda'}{\xi}{\tilde{\theta}_\psp} e^{i\xi\alpha},
\end{split}
\label{eq:ampz}
\end{equation}
where $\lambda'$ is the helicity of the $\psp$
(the quantization axis being parallel to the $\pim$ momentum in the $\psp$
rest frame);
$\theta_{\z}$ is the $\z$ helicity angle (the angle between the $\kp$ and
$\pim$ momenta in the $\z$ rest frame);
$\tilde{\varphi}$ is the angle between the planes defined by
the $(\lp,\pim)$ and $(\kp,\pim)$ momenta in the $\psp$ rest frame; 
$\tilde{\theta}_{\psp}$ is the $\psp$ helicity angle (the angle between the 
$\pim$ and $\mu^-$ momenta in the $\psp$ rest frame);
$\alpha$ is the angle between the planes defined by
the $(\lp,\pim)$ and $(\lp,\kst)$ momenta in the $\psp$ rest frame.
If the spin of the $\z$ equals 0, only $\lambda'=0$ is allowed.
The amplitudes in Eq.~\eqref{eq:ampz} for different $\lambda'$ values
are related by parity conservation:
\begin{equation}
 H_{\lambda'}^{\z} = -P(\z)(-1)^{J(\z)} H^{\z}_{-\lambda'}. \\
\end{equation}

The resulting expression for the signal density function is
\begin{equation}
\label{eq:sig_pdf}
\begin{split}
S(\Phi) = 
\sum_{\xi=1,-1} \left|
\sum_{\kst} \sum_{\lambda=-1,0,1} A_{\lambda\,\xi}^{\kst} +
\sum_{\lambda'=-1,0,1} A_{\lambda'\,\xi}^{\z}
\right|^2.
\end{split}
\end{equation}
A detailed description of the derivation of the amplitude is given in the
Appendix.

We perform an unbinned maximum likelihood fit in
the four-dimensional space $\Phi$.
The construction of the likelihood function follows Ref.~\cite{garmash}.
The function to be minimized is
\begin{equation}
\label{eq:fcn}
F = - 2 \sum\limits_{i}
\ln\Big( (1-b)\frac{S(\Phi_i)}
{\sum\limits_{j}S(\Phi_j)}
 +b\frac{B(\Phi_i)}{\sum\limits_{j}B(\Phi_j)}\Big),
\end{equation}
where $b$ is the fraction of the background events and $B(\Phi)$
is the background density in the signal region.
The sum $\sum\limits_{i}$ runs over data events;
the sum $\sum\limits_{j}$
runs over MC events generated uniformly over the phase space
and reconstructed using the same selection requirements as in data.
This procedure takes into account the nonuniformity of the
reconstruction efficiency but requires
a parametrization of the background shape.

As there is sensitivity only to the relative phases
of the various contributions, the phase of $H^{\kst(892)}_0$ is fixed to zero.
The detector resolution
in $M_{K \pi}$ and $M_{\psp \pi}$ ($\sigma\sim3\ \mevcc$) is small compared to
the width of any of the resonances that are considered and is ignored.
The masses and widths of all the $\kst$ resonances except $K^*_0(800)$
are fixed to their nominal values~\cite{PDG}.
The mass and width of $K^*_0(800)$ are fixed to the fit results
in the default model without a $\z$
(M = $946\pm50\ \mevcc$, $\Gamma$ = $736\pm126\ \mev$);
the case of free mass and width is included to systematic uncertainty.
We do not constrain the mass and the width of the $\zm{4430}$
to the previously measured values~\cite{z4430dalitz}.
The $r$ parameters in the Blatt-Weisskopf form factors
 are fixed at a default value of $1.6\ \gev^{-1}$.

\section{Results}
\label{sec:results}

\subsection{Fit to the background distribution}

\begin{figure}[ht]
\begin{center}
\includegraphics[width=4.2cm]{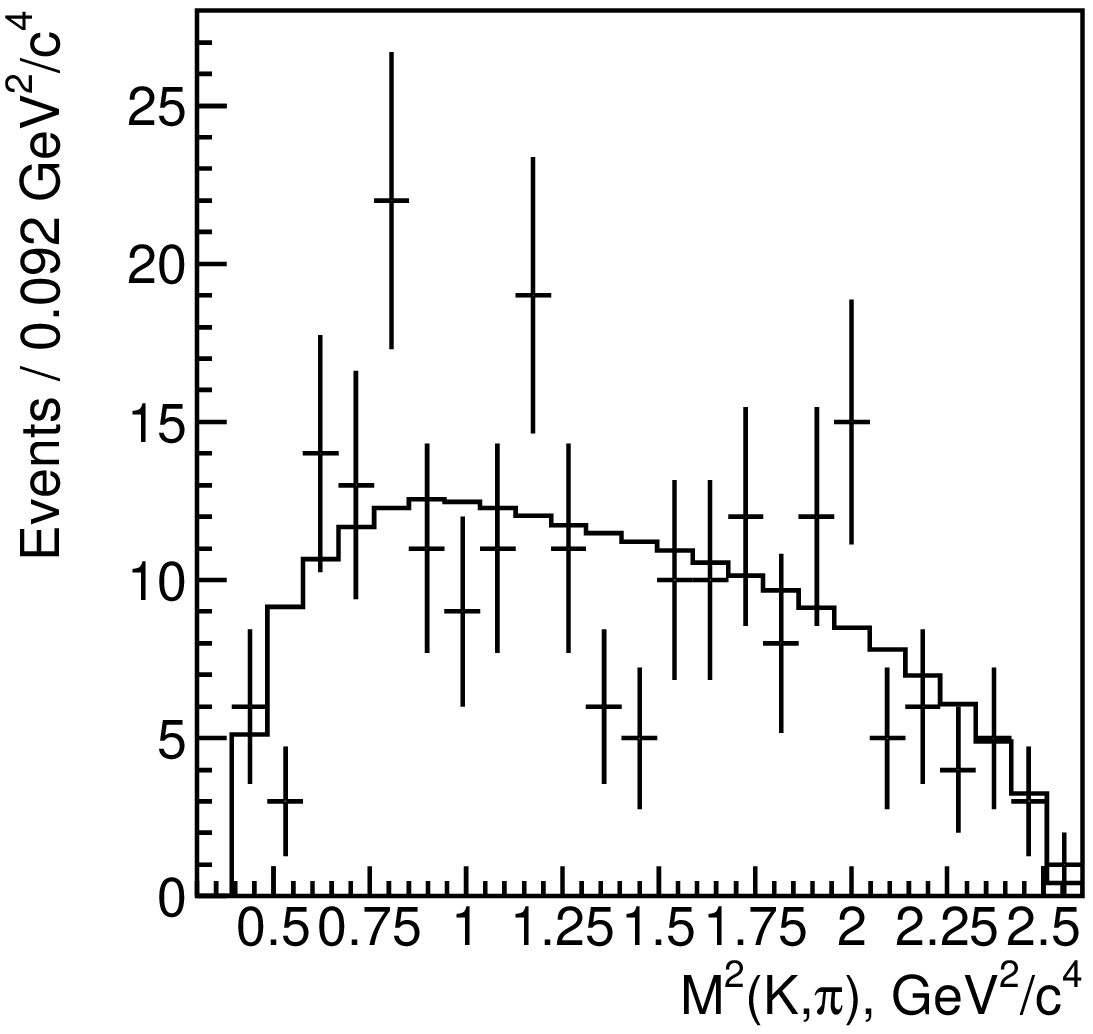}
\includegraphics[width=4.2cm]{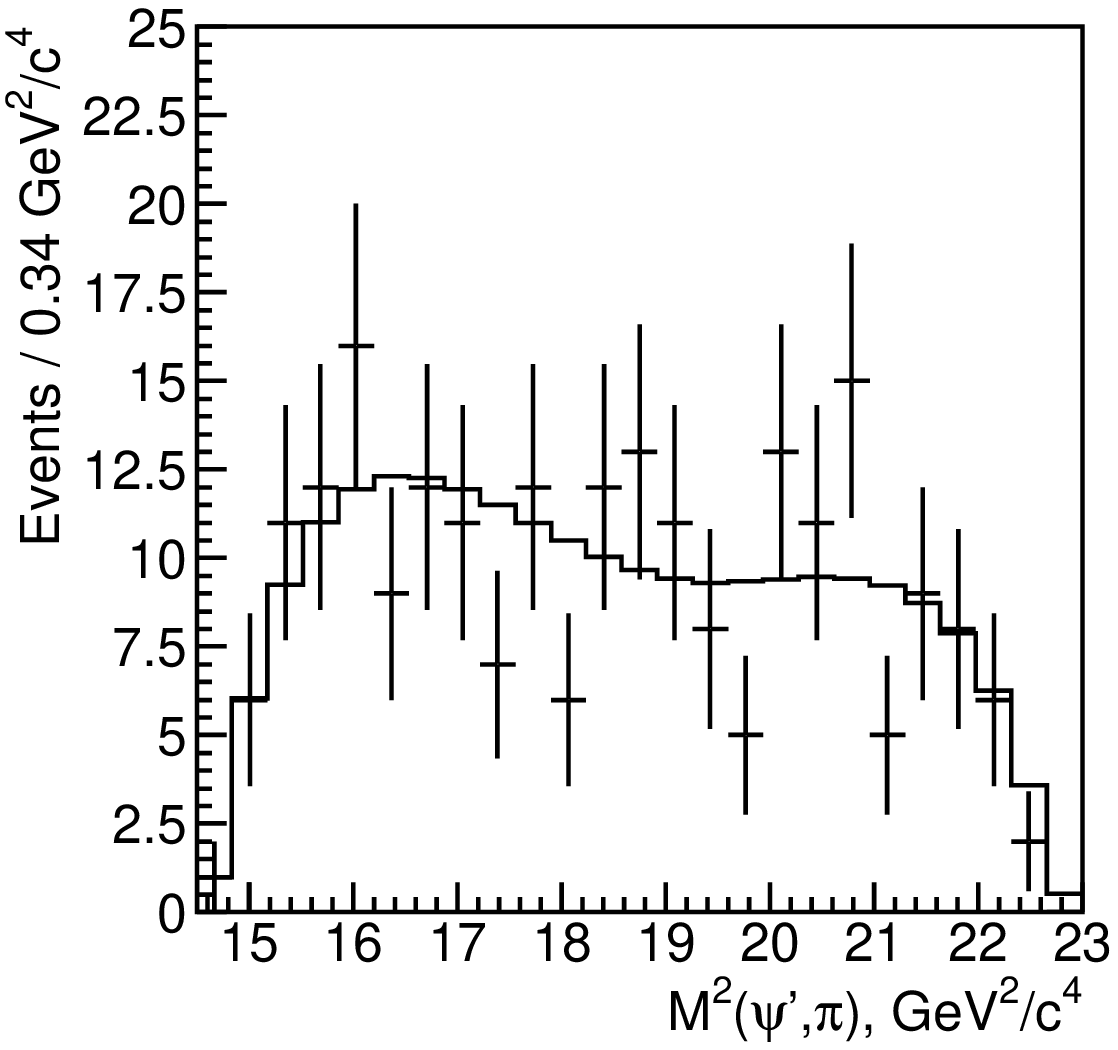}
\end{center}
\caption{The results of the fit to background events projected onto
the Dalitz variables.}
\label{fig:sbfitres}
\end{figure}

The background shape is determined using $\DE$ sidebands.
The background density function is defined as
\begin{equation}
B(\Phi) = P_2(\sx,\,\sy),
\label{eq:bgpdf}
\end{equation}
where $P_2$ is a two-dimensional second-order polynomial.
We perform an unbinned maximum likelihood fit; the function to be
minimized is given by Eq.~\eqref{eq:fcn} with $b=1$;
thus, the resulting $B(\Phi)$ is efficiency corrected.
The results of the fit, projected onto the Dalitz variables,
are shown in Fig.~\ref{fig:sbfitres}.
If the angular variables are also considered and 
the fitting function is multiplied by additional polynomials
$P^{(\varphi)}_2(\varphi)$ and
$P^{(\theta)}_2(\theta_{\psp})$,
the coefficients of the nonconstant terms are consistent with zero
after minimization;
thus, the background does not depend on the angular variables.

\subsection{Fit to the data}

The fit results for the $\zm{4430}$ mass, width and significance
in the default model are shown in
Table~\ref{tab:z_jp} for all spin-parity hypotheses
with $J\leq2$. Note that the $0^+$ assignment is forbidden by
parity conservation in $\zm{4430}\to\psp\pim$ decays.
The significance of the $\zm{4430}$ is estimated from the difference
of $-2\ln L$ between the models with and without a $\zm{4430}$ signal,
taking into account the number of added degrees of freedom
(6 for the $1^+$ and $2^-$ hypotheses or 4 for other hypotheses).
The preferred $\zm{4430}$ spin-parity hypothesis is $1^+$.
To test the goodness of fit we bin the Dalitz distribution with the requirement
that the number of events in each bin $n_i>16$. We then calculate the $\chi^2$
value as $\sum_i (n_i-s_i)^2/s_i$, where $s_i$ is the integral of the fitting
function over the bin $i$. We generate MC pseudoexperiments in accordance with
the result of the fit; the confidence level is defined as the fraction of the
pseudoexperiments with the $\chi^2$ value greater than the $\chi^2$ value
in data. The confidence level of the $1^+$ hypothesis is 15\%.
The absolute values and phases of the amplitudes for the $1^+$ hypothesis
are listed in Table~\ref{tab:kstamp}.
The significances of the $K^*$ resonances are shown in Table~\ref{tab:ffrac}.

\begin{table*}[ht]
\caption{
Fit results in the default model. Errors are statistical only.
}
\begin{tabular}{c|c|c|c|c|c} 
\hline\hline
$J^P$ & $0^-$ & $1^-$ & $1^+$ & $2^-$ & $2^+$ \\
\hline
Mass, $\mevcc$ & $4479\pm16$ & $4477\pm4$ & $4485\pm20$ & $4478\pm22$ & $4384\pm19$ \\
Width, $\mev$ & $110\pm50$ & $22\pm14$ & $200\pm40$ & $83\pm25$ & $52\pm28$ \\
Significance & $4.5\sigma$ & $3.6\sigma$ & $6.4\sigma$ & $2.2\sigma$ & $1.8\sigma$ \\
\hline\hline
\end{tabular}
\label{tab:z_jp}
\end{table*}

\begin{table*}
\caption{The absolute values and phases of the amplitudes in the default model for
the $1^+$ spin-parity of the $\zm{4430}$. Errors are statistical only.}
\begin{tabular}{c|c|c|c|c|c|c}
\hline\hline
Resonance & $a_0$ & $\phi_0$ & $a_1$ & $\phi_1$ & $a_{-1}$ & $\phi_{-1}$ \\
\hline
$K^*_0(800)$  & $2.03\pm0.44$ & $1.87\pm0.22$ & $\ldots$ & $\ldots$ & $\ldots$ & $\ldots$ \\ 
$K^*(892)$    & $1$ (fixed) & $0$ (fixed) & $0.81\pm0.07$ & $-2.79\pm0.12$ & $0.43\pm0.08$ & $-1.64\pm0.15$ \\
$K^*(1410)$   & $0.52\pm0.22$ & $0.12\pm0.66$ & $0.47\pm0.44$ & $-1.38\pm0.55$ & $0.57\pm0.31$ & $1.38\pm0.66$ \\
$K^*_0(1430)$ & $1.08\pm0.50$ & $-2.57\pm0.63$ & $\ldots$ & $\ldots$ & $\ldots$ & $\ldots$ \\
$K^*_2(1430)$ & $8.48\pm2.45$ & $-0.41\pm0.33$ & $12.6\pm4.2$ & $2.56\pm0.69$ & $6.44\pm4.21$ & $-2.44\pm1.12$ \\
$K^*(1680)$   & $0.31\pm0.51$ & $2.08\pm0.17$ & $1.91\pm0.77$ & $3.08\pm0.26$ & $0.48\pm0.59$ & $-1.94\pm2.03$ \\
$\zm{4430}$   & $8.85\pm2.57$ & $-2.97\pm0.77$ & $8.83\pm2.75$ & $-2.80\pm0.27$ & \multicolumn{2}{c}{$(a_{-1}e^{i\phi_{-1}}) = (a_{1}e^{i\phi_{1}}$)} \\
\hline\hline
\end{tabular}
\label{tab:kstamp}
\end{table*}

\begin{figure}[ht]
\begin{center}
\includegraphics[width=6cm]{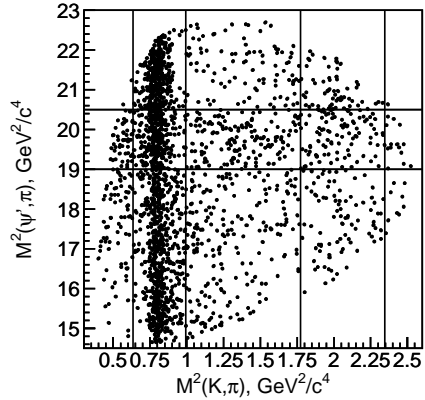}
\end{center}
\caption{Dalitz plot slices used to present fit results. Vertical
divisions are at $(0.796)^2\ \gevccsq$, $(0.996)^2\ \gevccsq$,
$(1.332)^2\ \gevccsq$ and $(1.532)^2\ \gevccsq$.
Horizontal divisions are at $19.0\ \gevccsq$ and $20.5\ \gevccsq$.}
\label{fig:slices}
\end{figure}

\begin{figure*}[ht]
\begin{center}
\includegraphics[width=4.2cm,height=4.2cm]{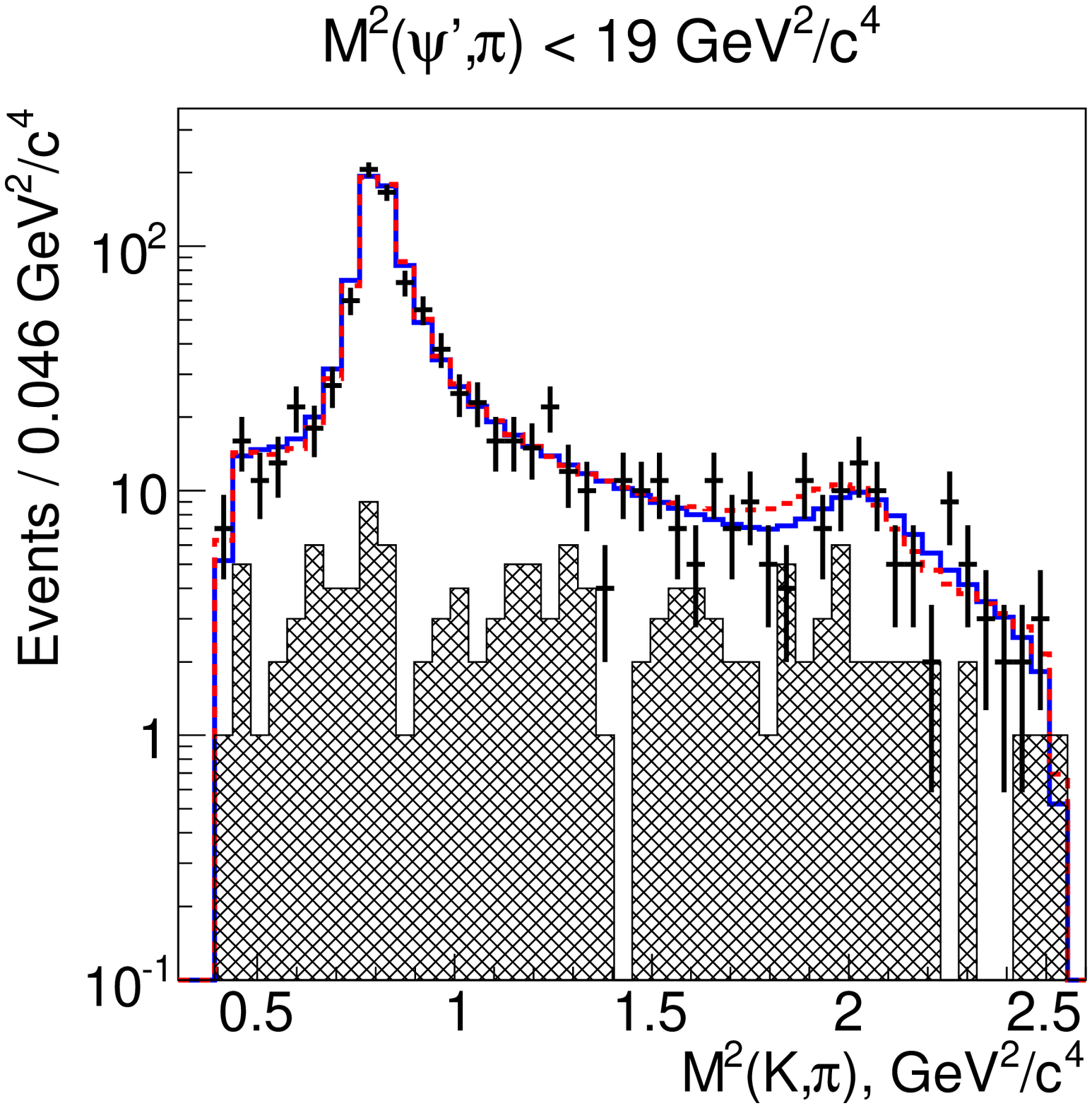} 
\includegraphics[width=4.2cm,height=4.2cm]{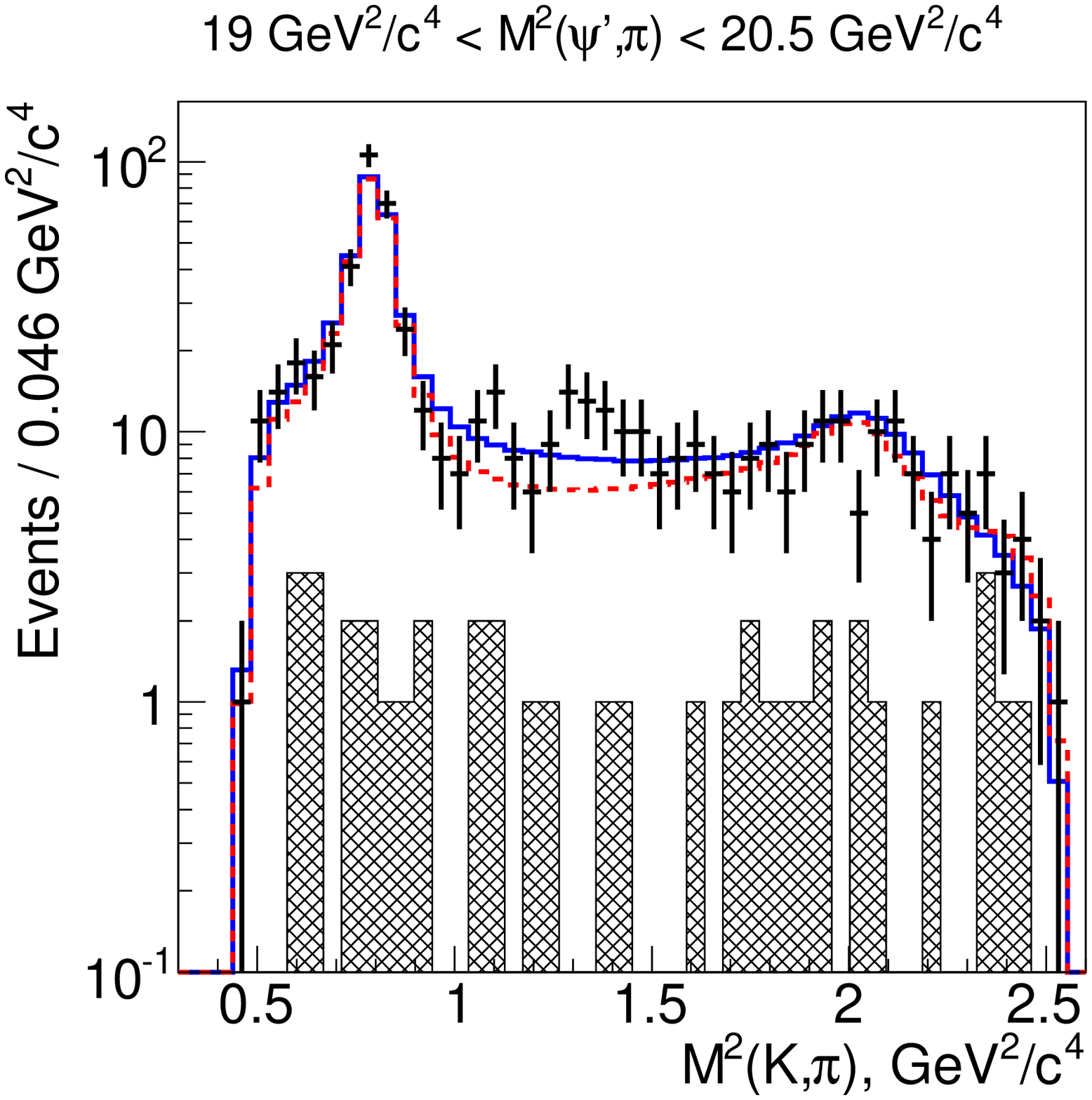} 
\includegraphics[width=4.2cm,height=4.2cm]{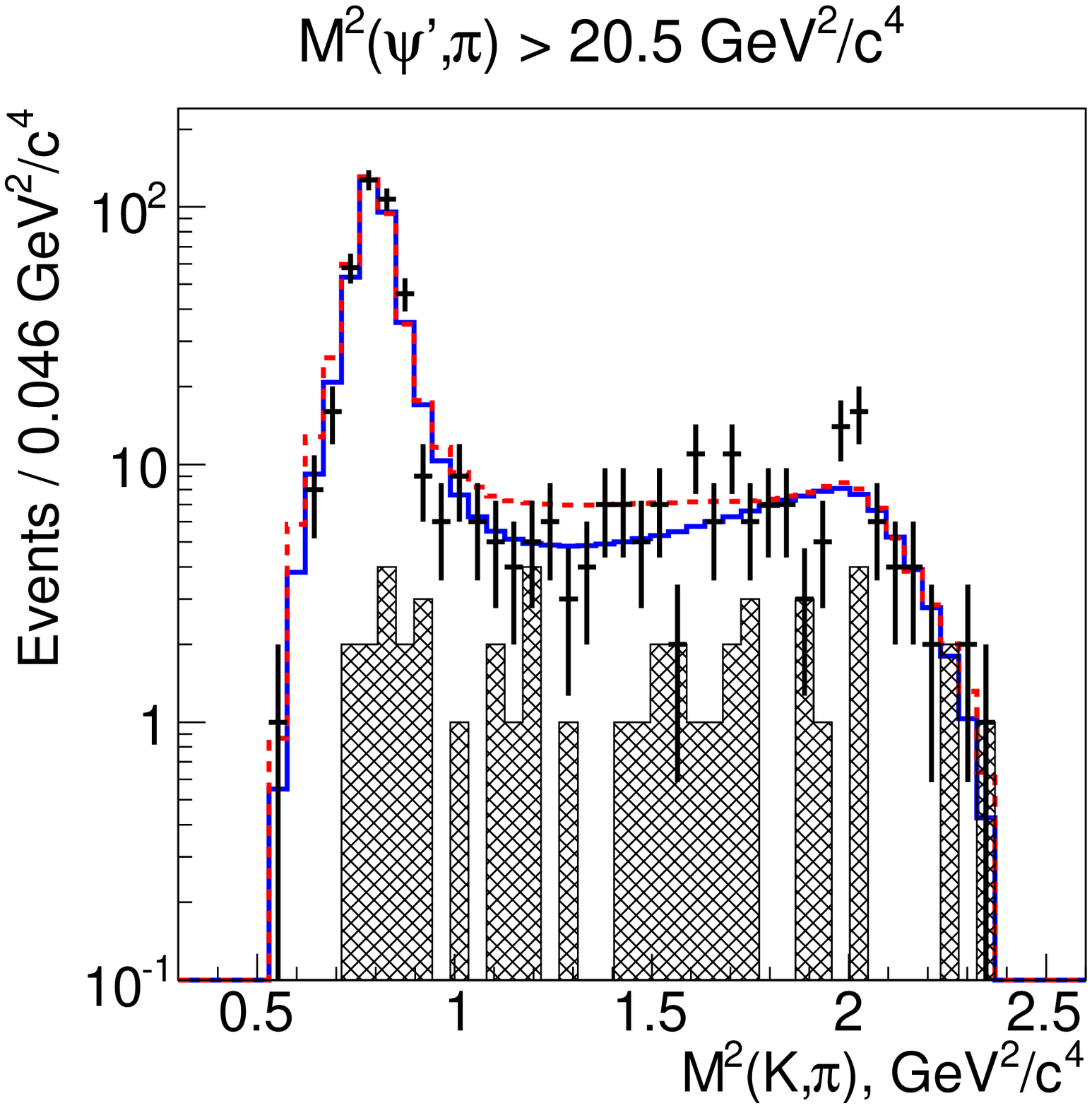} 
\includegraphics[width=4.2cm,height=4.2cm]{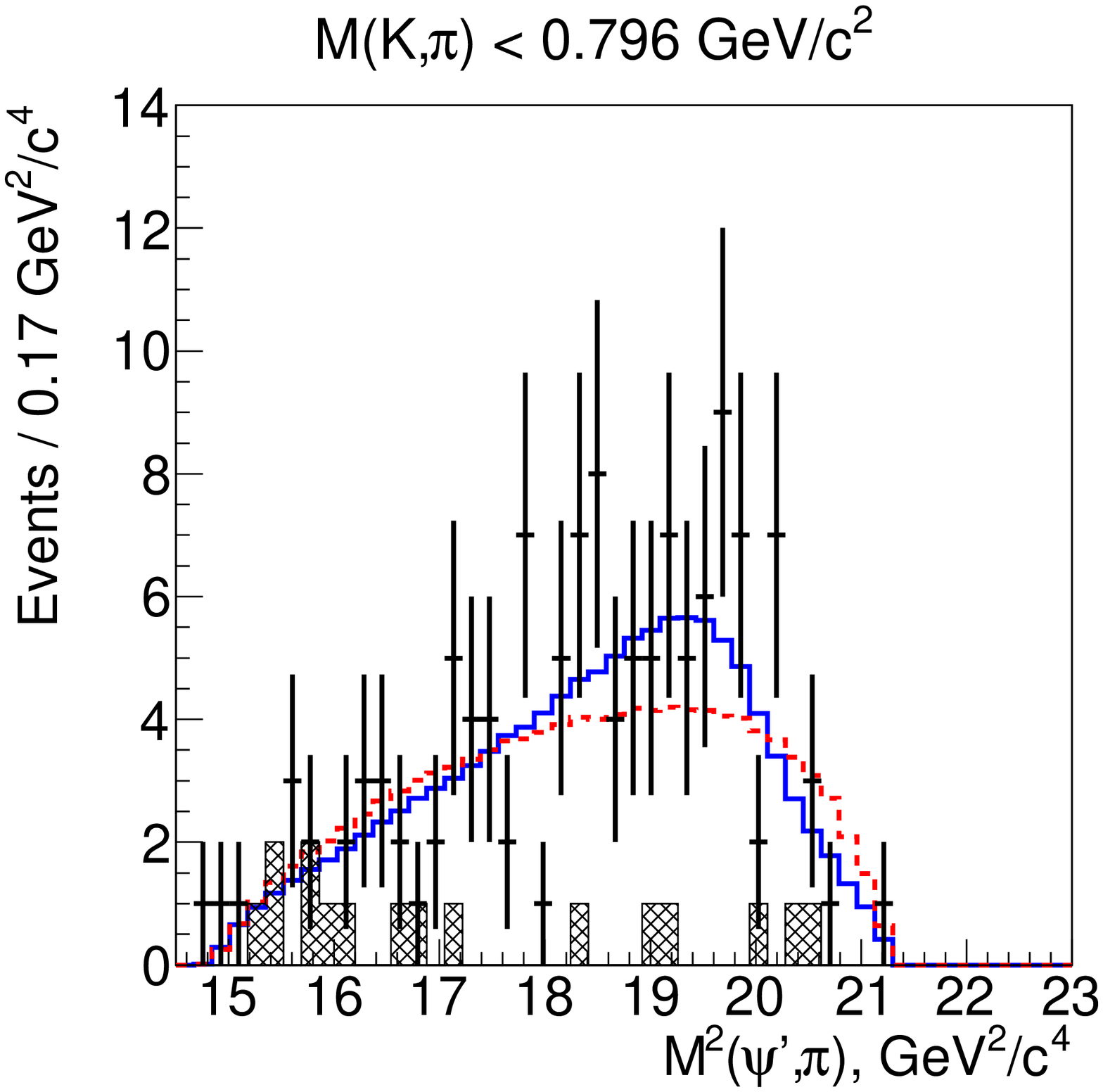} 
\includegraphics[width=4.2cm,height=4.2cm]{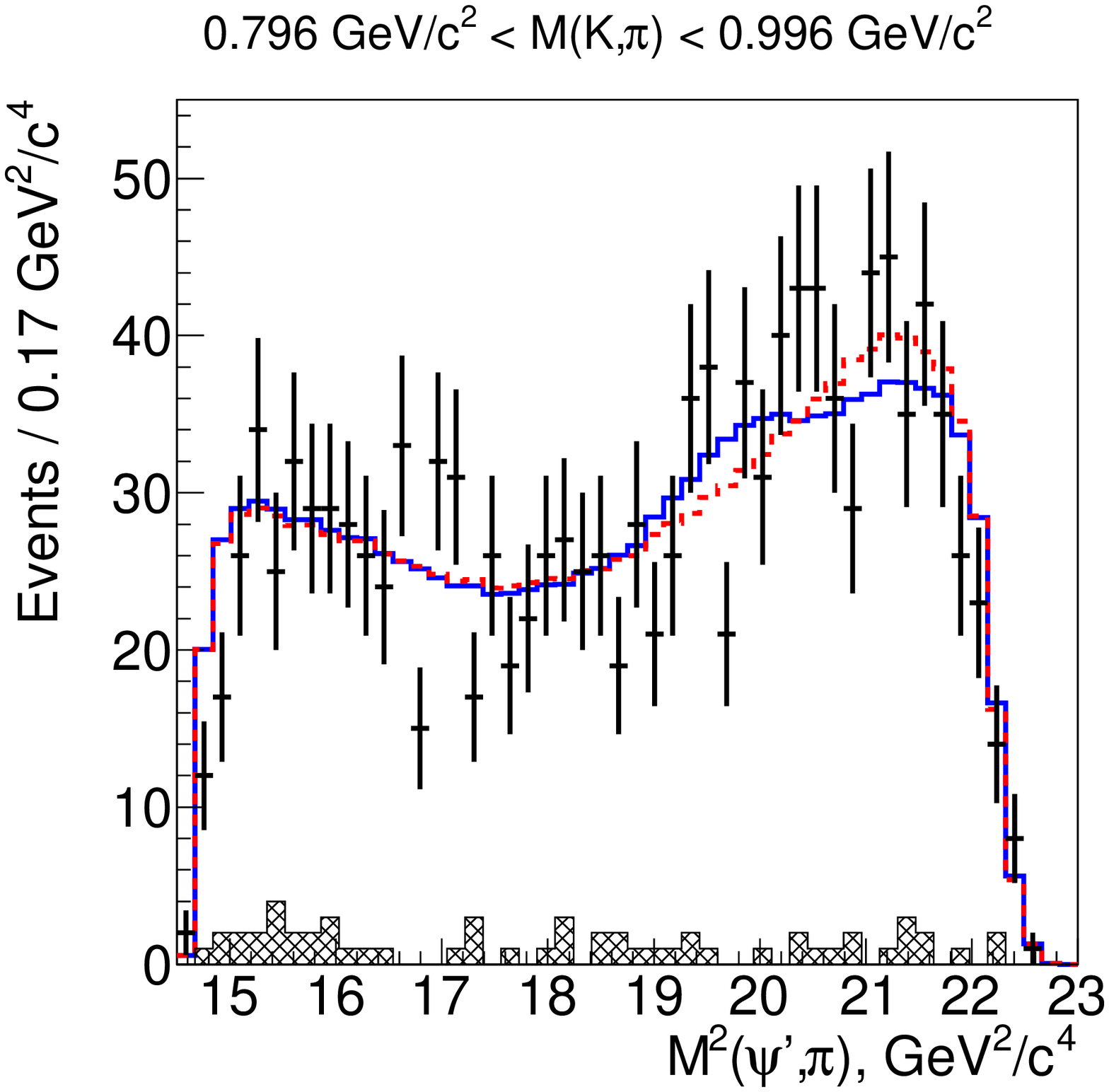} 
\includegraphics[width=4.2cm,height=4.2cm]{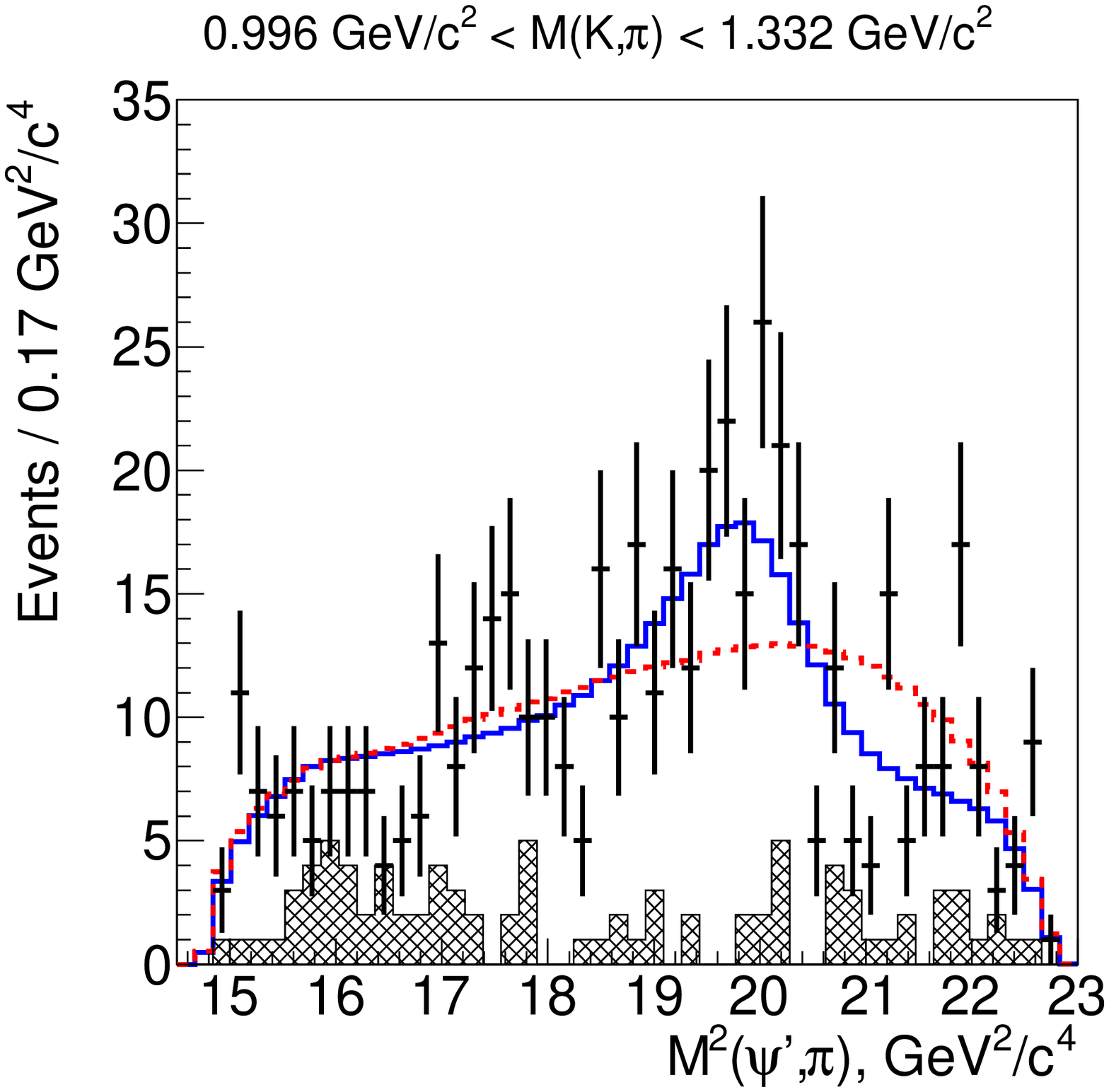} 
\includegraphics[width=4.2cm,height=4.2cm]{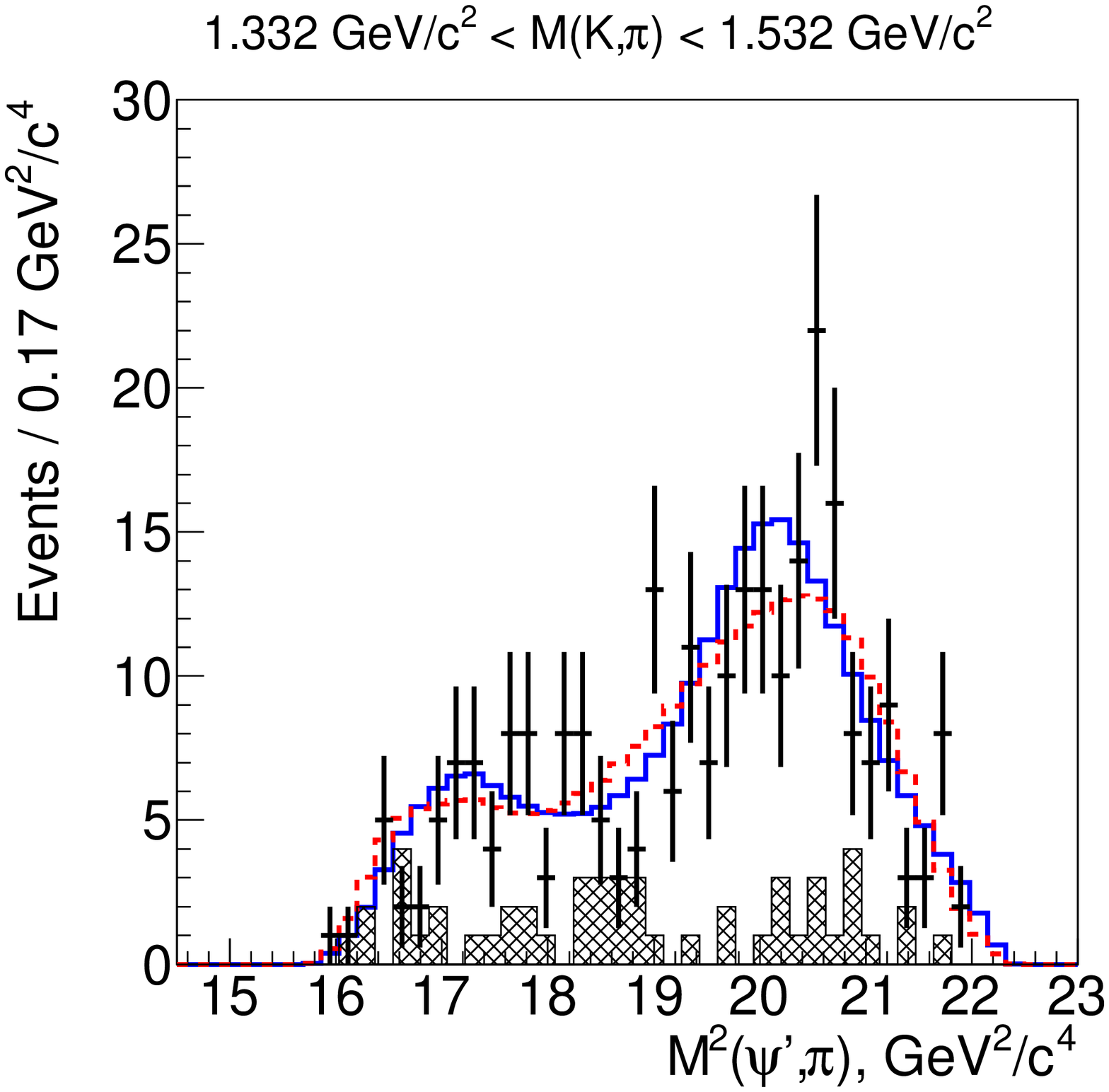} 
\includegraphics[width=4.2cm,height=4.2cm]{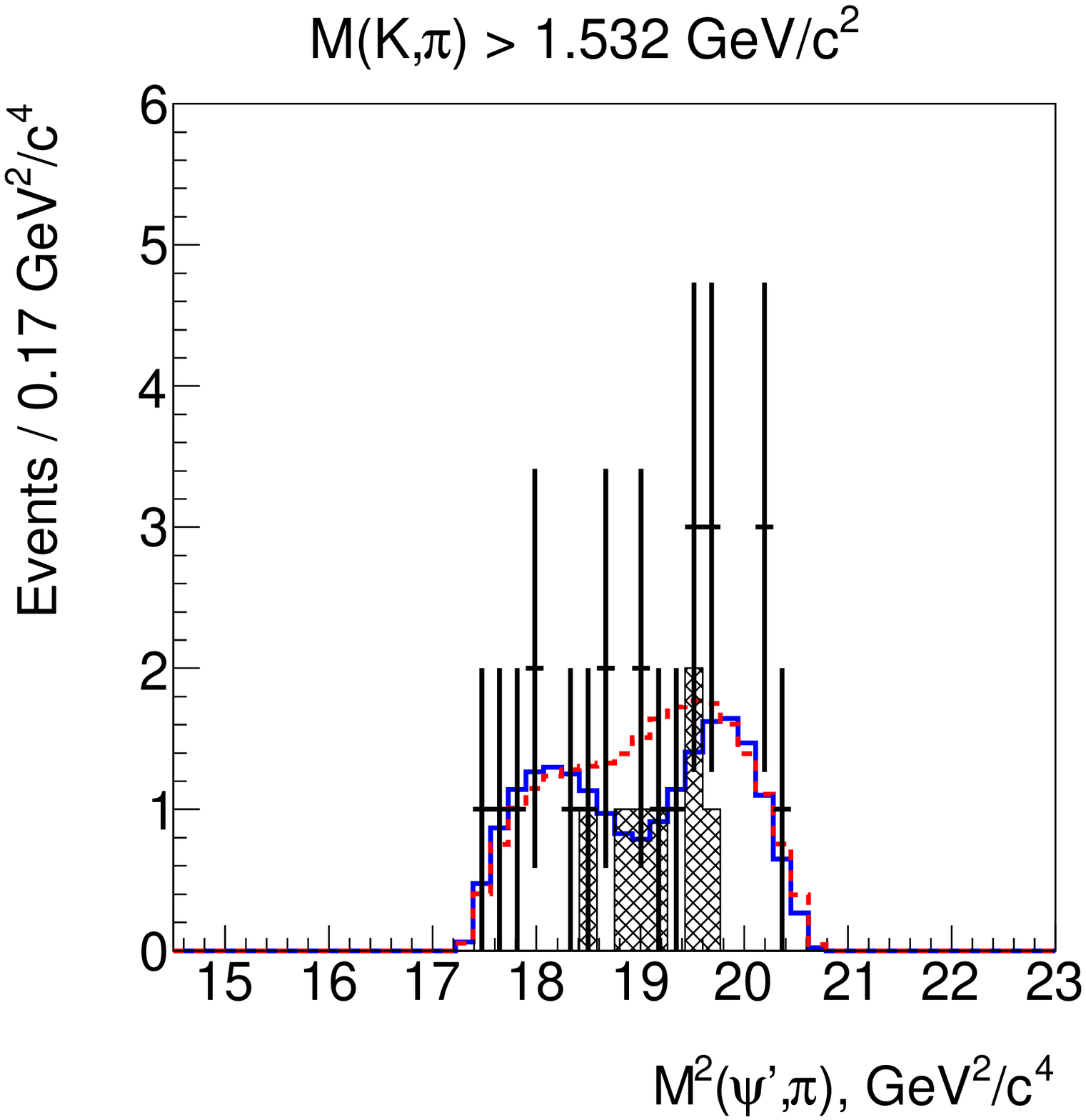} 
\end{center}
\caption{The fit results with (solid line) and without (dashed line)
$\z$ ($J^P=1^+$) in the default model. The points with error bars are data;
the hatched histograms are $\psp$ sidebands. Slices are defined in
Fig.~\ref{fig:slices}.}
\label{fig:fitresdef}
\end{figure*}

\begin{figure}
\includegraphics[width=6cm]{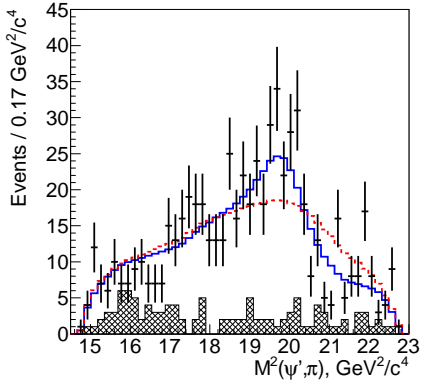}
\caption{Projection of the fit results with the $K^*$ veto.
The legend is the same as in Fig.~\ref{fig:fitresdef}.}
\label{fig:zveto}
\end{figure}

\begin{figure}
\begin{center}
\includegraphics[width=6cm]{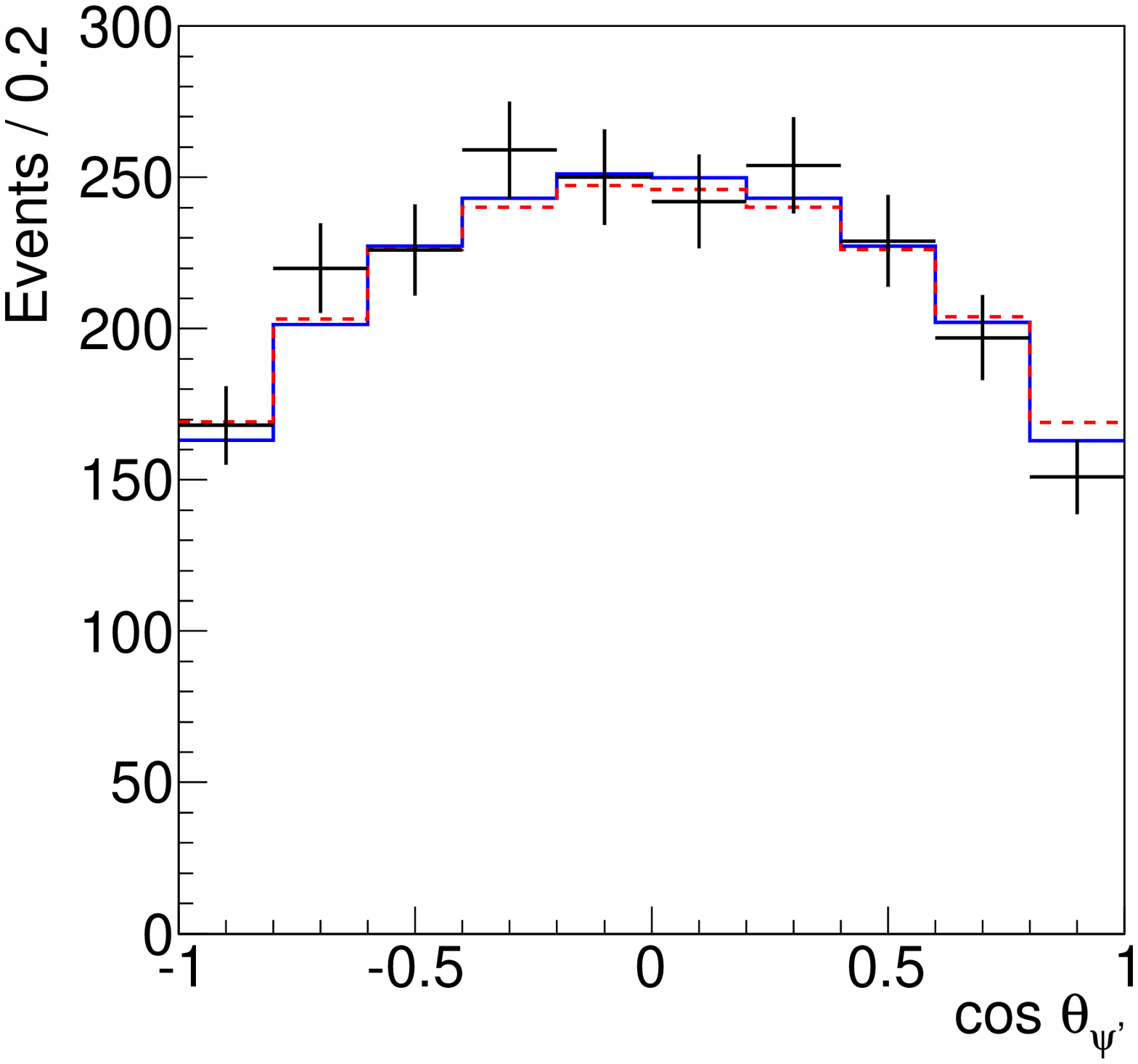}
\includegraphics[width=6cm]{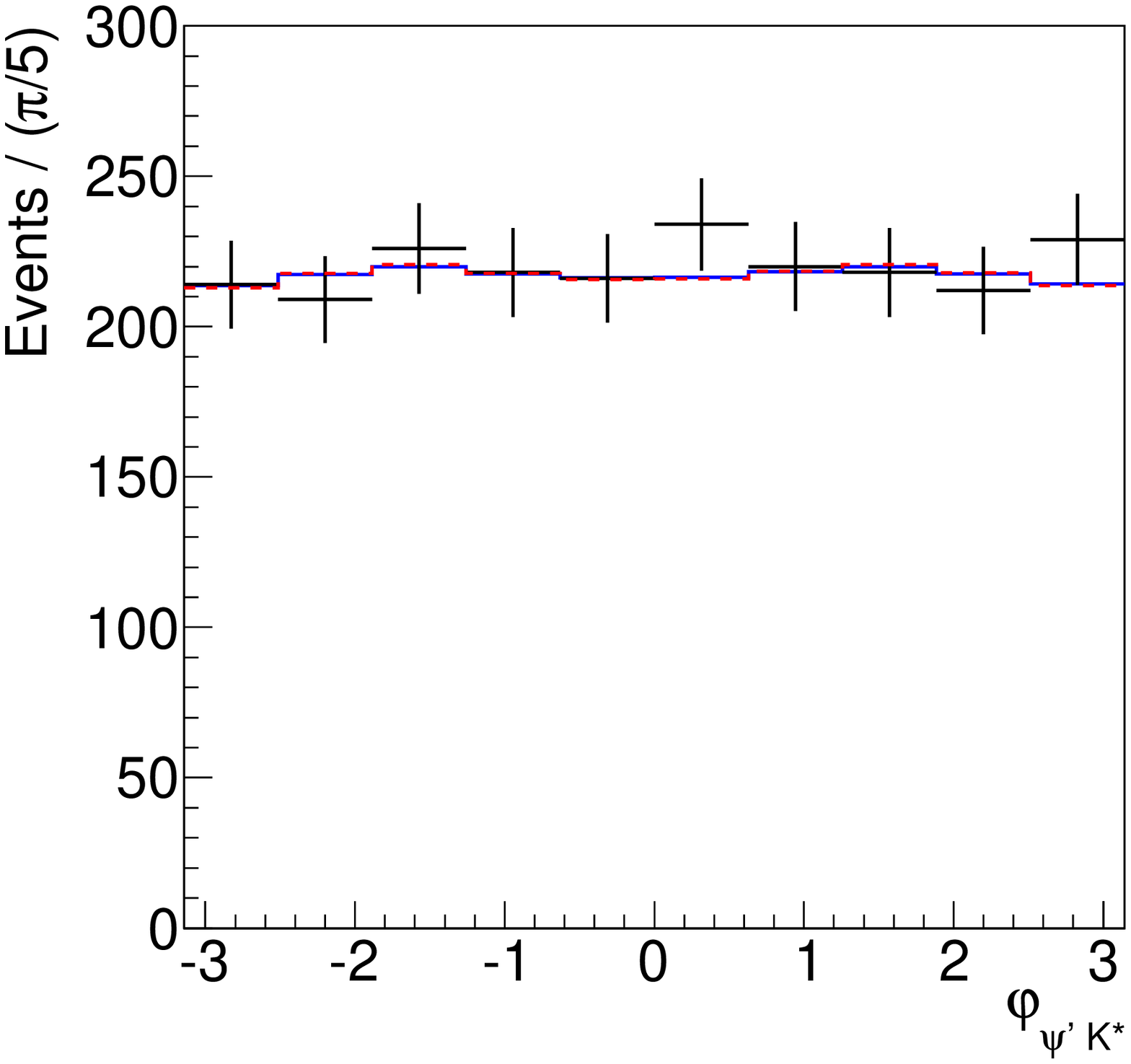}
\end{center}
\caption{Projections of the fit results with (solid line) and without
(dashed line) $\z$ ($J^P=1^+$) onto the angular variables for the entire signal
region (no $\kst$ veto) in the default model. Points with error bars are data.}
\label{fig:fitresang}
\end{figure}

To present the fit results, the Dalitz distribution is divided into slices
that are shown in Fig.~\ref{fig:slices}. The second and fourth vertical slices
correspond to the regions of the $K^*(892)$ and $K^*_2(1430)$, respectively;
the second horizontal slice corresponds to the region of the $\zm{4430}$.
%
Projections of the fit results
onto $M^2_{K \pi}$ and $M^2_{\psp \pi}$ axes for the $1^+$ hypothesis and the
model without $\zm{4430}$
are shown in Fig.~\ref{fig:fitresdef}.
The sum of the first, third and fifth vertical slices
[the $M^2(\psp \pi)$ projection with the $\kst$ veto applied]
is shown in Fig.~\ref{fig:zveto}. Projections onto the
angular variables are shown in Fig.~\ref{fig:fitresang}.

\begin{table}
\caption{The fit fractions and significances of all resonances in the default
model ($J^P=1^+$).}
\begin{tabular}{c|c|c}
\hline\hline
Resonance & Fit fraction & Significance \\
\hline
$K^*_0(800)$  &
$(5.8\pm2.1)\%$  & $3.6\sigma$  \\
$K^*(892)$    &
$(63.8\pm2.6)\%$ & $43.1\sigma$ \\
$K^*(1410)$   &
$(4.3\pm2.3)\%$  & $0.6\sigma$  \\
$K^*_0(1430)$ &
$(1.1\pm1.4)\%$  & $1.6\sigma$  \\
$K^*_2(1430)$ &
$(4.5\pm1.0)\%$  & $3.3\sigma$  \\
$K^*(1680)$   &
$(4.4\pm1.9)\%$  & $1.0\sigma$  \\
$\zm{4430}$   &  $(10.3^{+3.0}_{-3.5})\%$ & $6.4\sigma$ \\
\hline\hline
\end{tabular}
\label{tab:ffrac}
\end{table}

We also consider other amplitude models, including models without one
of the insignificant $\kst$ resonances
[$\kst(1410)$, $\kst_0(1430)$, $\kst(1680)$];
with the addition of nonresonant $\kp\pim$ amplitudes in S, P and D waves;
with free Blatt-Weisskopf $r$ parameters;
with free masses and widths of $K^*$ resonances
(within their uncertainties~\cite{PDG}) and with LASS
amplitudes~\cite{LASS} instead of Breit-Wigner amplitudes for all spin-0 $K^*$
resonances.

In Ref.~\cite{z4430dalitz}, the assigned value of the $B$-decay
orbital angular momentum ($L$) is varied to study the systematic
uncertainty. In this analysis, we instead change the default helicity amplitudes
with minimal $L$ to partial wave amplitudes with known $L$
(which does not result in a significant improvement of the likelihood);
this model is also included in the systematic uncertainty.

The significances of the $\zm{4430}$ for all models other than the default one
are shown in Table~\ref{tab:zsigsyst}.
The significance of the $1^+$ hypothesis is above
$5.0\sigma$ for all the models.

\begin{table}
\caption{Model dependence of the $\zm{4430}$ significance.}
\begin{tabular}{c|c|c|c|c|c}
\hline\hline
Model & $0^-$ & $1^-$ & $1^+$ & $2^-$ & $2^+$ \\
\hline
Without $K^*(1410)$        &
$3.8\sigma$ & $3.4\sigma$ & $6.9\sigma$ & $2.1\sigma$ & $1.0\sigma$ \\
Without $K^*_0(1430)$        &
$4.9\sigma$ & $3.5\sigma$ & $7.4\sigma$ & $1.4\sigma$ & $1.0\sigma$ \\
Without $K^*(1680)$        &
$4.2\sigma$ & $3.3\sigma$ & $7.2\sigma$ & $2.6\sigma$ & $1.4\sigma$ \\
With $K^*_3(1780)$         &
$2.9\sigma$ & $3.1\sigma$ & $5.2\sigma$ & $2.2\sigma$ & $1.6\sigma$ \\
LASS                       &
$4.3\sigma$ & $3.5\sigma$ & $6.2\sigma$ & $2.9\sigma$ & $1.6\sigma$ \\
Partial wave amplitudes    &
$4.6\sigma$ & $3.5\sigma$ & $6.8\sigma$ & $2.4\sigma$ & $1.8\sigma$ \\
Free masses and widths     &
$4.8\sigma$ & $3.5\sigma$ & $6.4\sigma$ & $2.7\sigma$ & $2.0\sigma$ \\
Free $r$                   &
$4.1\sigma$ & $3.7\sigma$ & $6.4\sigma$ & $2.4\sigma$ & $1.9\sigma$ \\
Nonresonant ampl. (S)     &
$5.1\sigma$ & $3.6\sigma$ & $6.8\sigma$ & $2.7\sigma$ & $1.7\sigma$ \\
Nonresonant ampl. (S,P)   &
$5.4\sigma$ & $3.6\sigma$ & $6.9\sigma$ & $3.0\sigma$ & $2.2\sigma$ \\
Nonresonant ampl. (S,P,D) &
$3.6\sigma$ & $2.7\sigma$ & $5.6\sigma$ & $2.2\sigma$ & $1.4\sigma$ \\
\hline\hline
\end{tabular}
\label{tab:zsigsyst}
\end{table}

The exclusion levels of the spin-parity hypotheses ($J^P=j^p$)
are calculated from MC simulation. For each amplitude model,
we generate MC pseudoexperiments in accordance with the fit result
with the $j^p$ $\zm{4430}$ signal in data and fit them with
the $j^p$ and $1^+$ signals.
The resulting distribution of
$\Delta(-2\ln L) = (-2\ln L)_{J^P=j^p} - (-2\ln L)_{J^P=1^+}$
is fitted to an asymmetrical Gaussian function and the $p$-value
is calculated as the integral of the fitting function normalized to 1
from the value of
$\Delta(-2\ln L)$ in data to $+\infty$. The results are presented in Table
~\ref{tab:excl}.
The $J^P=1^+$ hypothesis is favored over the $0^-$, $1^-$, $2^-$ and $2^+$
hypotheses at the levels of $3.4\sigma$, $3.7\sigma$, $4.7\sigma$ and
$5.1\sigma$, respectively.

\begin{figure}
\includegraphics[width=6cm]{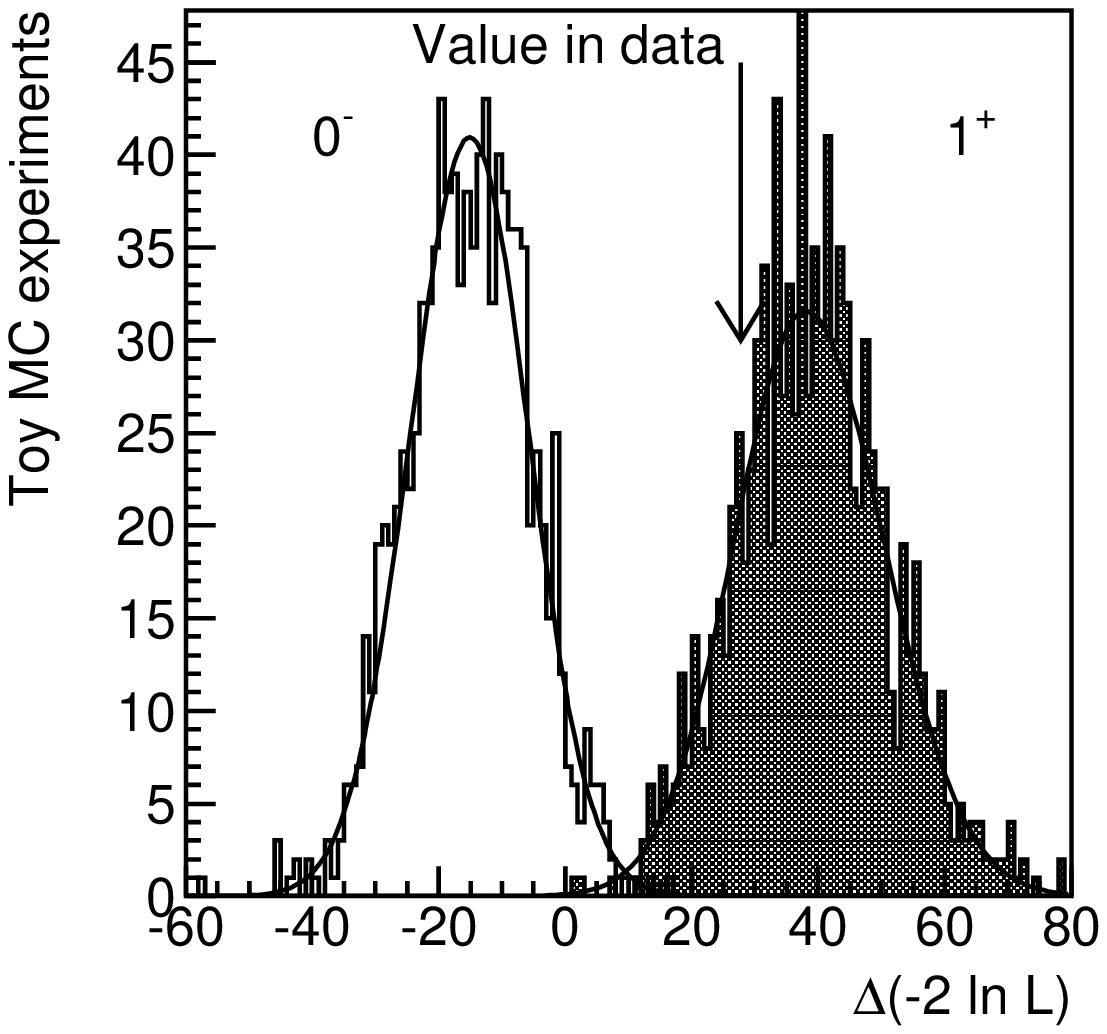}
\caption{Comparison of the $0^-$ and $1^+$ hypotheses in the default model.
The histograms are distributions of $\Delta(-2\ln L)$ in MC pseudoexperiments
generated in accordance with the fit results with $0^-$
(open histogram) and $1^+$ (hatched histogram) signals.
The $\Delta(-2\ln L)$ value observed in data is indicated with an arrow.}
\label{fig:toy}
\end{figure}

We also generate MC pseudoexperiments in accordance with the fit results for
the $1^+$ hypothesis and obtain the distribution of $\Delta(-2\ln L)$.
This distribution is fitted to an asymmetrical Gaussian function and the 
confidence level of the $1^+$ hypothesis is calculated as the integral
of the fitting function
normalized to 1 from $-\infty$ to the value of $\Delta(-2\ln L)$ in data.
The resulting confidence levels are shown in Table~\ref{tab:excl}.
The distributions of $\Delta(-2\ln L)$ for $j^p=0^-$
are shown in Fig.~\ref{fig:toy}.

The results of the study of the model dependence of the $\zm{4430}$
mass and width are shown in Table~\ref{tab:syst}.
The maximal deviations of the mass and the width of the $\zm{4430}$
from their optimal values are considered as overall systematic uncertainty
due to the amplitude model dependence. We also estimate the uncertainty due
to the uncertainties of the fit to the background distribution by varying
the background parameters by $\pm1\sigma$ (with other parameters varied in
accordance with the correlation coefficients)
and performing the fit to the data.
The maximal deviations are considered as systematic uncertainty.

\begin{table}
\caption{Systematic uncertainties in the $\zm{4430}$ mass (in $\mevcc$)
and width (in $\mev$).}
\begin{tabular}{c|c|c}
\hline\hline
Model or error source & $\ $Mass $\ $ & Width \\
\hline
Without $K^*(1410)$        &
$^{+4}_{-0}$  & $^{+0}_{-9}$  \\
Without $K^*_0(1430)$        &
$^{+18}_{-0}$  & $^{+24}_{-0}$  \\
Without $K^*(1680)$        &
$^{+27}_{-0}$  & $^{+0}_{-32}$ \\
With $K^*_3(1780)$         &
$^{+5}_{-0}$ & $^{+23}_{-0}$  \\
LASS                       &
$^{+0}_{-3}$  & $^{+13}_{-0}$ \\
Partial wave amplitudes    &
$^{+12}_{-0}$  & $^{+0}_{-26}$ \\
Free masses and widths     &
$^{+0}_{-1}$  & $^{+0}_{-4}$  \\
Free $r$                   &
$^{+13}_{-0}$ & $^{+9}_{-0}$ \\
Nonresonant ampl. (S)     &
$^{+0}_{-9}$  & $^{+13}_{-0}$  \\
Nonresonant ampl. (S,P)   &
$^{+0}_{-11}$  & $^{+8}_{-0}$  \\
Nonresonant ampl. (S,P,D) &
$^{+2}_{-0}$  & $^{+9}_{-0}$  \\
\hline
Amplitude model, total &
$^{+27}_{-11}$ & $^{+24}_{-32}$ \\
Background &
$^{+2}_{-1}$ & $^{+3}_{-9}$ \\
\hline
Total &
$^{+27}_{-11}$ & $^{+24}_{-33}$ \\
\hline\hline
\end{tabular}
\label{tab:syst}
\end{table}

\begin{table*}
\caption{Exclusion levels of spin-parity hypotheses and confidence levels of the $1^+$
hypothesis.}
\begin{tabular}{c|c|c|c|c|c|c|c|c}
\hline\hline
\multirow{2}{*}{Model} & \multicolumn{2}{|c|}{$0^-$} & 
\multicolumn{2}{|c|}{$1^-$} & \multicolumn{2}{|c|}{$2^-$} &
\multicolumn{2}{|c}{$2^+$} \\
\cline{2-9}
 & $1^+$ over $0^-$ & $1^+$ C.L. & $1^+$ over $1^-$ & $1^+$ C.L. & $1^+$ over $2^-$ & $1^+$ C.L. & $1^+$ over $2^+$ & $1^+$ C.L. \\
\hline
Default                    &
$4.7\sigma$ & 17\% &
$6.3\sigma$ & 16\% &
$6.5\sigma$ & 50\% &
$8.2\sigma$ & 38\% \\
Without $K^*(1410)$        &
$6.4\sigma$ & 40\% &
$7.2\sigma$ & 25\% &
$7.7\sigma$ & 43\% &
$9.2\sigma$ & 50\% \\
Without $K^*_0(1430)$        &
$5.0\sigma$ & 22\% &
$4.1\sigma$ & 19\% &
$8.9\sigma$ & 69\% &
$8.9\sigma$ & 33\% \\
Without $K^*(1680)$        &
$7.1\sigma$ & 54\%  &
$8.2\sigma$ & 58\% &
$10.0\sigma$ & 79\% &
$11.1\sigma$ & 75\% \\
With $K^*_3(1780)$         &
$3.4\sigma$ & 53\% &
$3.7\sigma$ & 9.8\% &
$4.7\sigma$ & 27\% &
$5.1\sigma$ & 29\% \\
LASS                       &
$4.8\sigma$ & 9.7\%  &
$6.3\sigma$ & 12\% &
$5.5\sigma$ & 28\% &
$8.2\sigma$ & 30\% \\
Partial wave amplitudes    &
$5.1\sigma$ & 30\%  &
$6.6\sigma$ & 28\% &
$7.6\sigma$ & 52\% &
$9.7\sigma$ & 46\% \\
Free masses and widths     &
$4.8\sigma$ & 15\% &
$6.0\sigma$ & 14\% &
$6.3\sigma$ & 37\% &
$7.4\sigma$ & 35\% \\
Free $r$                   &
$5.5\sigma$ & 19\%  &
$5.7\sigma$ & 26\% &
$6.5\sigma$ & 37\% &
$7.3\sigma$ & 43\% \\
Nonresonant ampl. (S)     &
$3.9\sigma$ & 18\% &
$5.0\sigma$ & 9.3\% &
$6.1\sigma$ & 38\% &
$8.4\sigma$ & 25\% \\
Nonresonant ampl. (S,P)   &
$3.4\sigma$ & 20\% &
$5.0\sigma$ & 18\% &
$6.2\sigma$ & 46\% &
$6.2\sigma$ & 34\% \\
Nonresonant ampl. (S,P,D) &
$3.8\sigma$ & 20\% &
$4.8\sigma$ & 14\% &
$5.2\sigma$ & 41\% &
$5.2\sigma$ & 26\% \\
\hline\hline
\end{tabular}
\label{tab:excl}
\end{table*}

\subsection{Efficiency and branching fractions}

We use the signal density function determined from the fits to
calculate the efficiency
\begin{equation}
\epsilon_0 = \frac
{\int S(\Phi) \epsilon(\Phi) d\Phi}
{\int S(\Phi) d\Phi},
\end{equation}
where $\epsilon(\Phi)$ is the phase-space-dependent efficiency.
The reconstruction efficiency is found to be $(28.3\pm1.2)\%$.
The central value is calculated for the default model with $\z$ ($J^P=1^+$).
The error includes
the uncertainty in track reconstruction efficiency ($1.4\%$),
the error from the particle identification efficiency difference
between MC and data ($3.8\%$) and the uncertainty due to the amplitude model
dependence ($0.5\%$).
The error due to MC statistics is negligibly small.
The efficiency includes the correction for the difference between the particle
identification efficiency in MC and data, $(94.2\pm3.5)\%$.

Using the obtained efficiency and the branching fractions of $\psp$ decays
to $\ee$ and $\mumu$~\cite{PDG}, we determine:
\begin{equation*}
\begin{split}
\br(&\decay)\times\br(\psill) = \\
&\qquad(9.12\pm0.30\pm0.51)\times10^{-6} \\
\end{split}
\end{equation*}
and
\begin{equation*}
\begin{split}
&\br(\decay) = (5.90\pm0.20\pm0.36)\times10^{-4}.
\end{split}
\end{equation*}
This result assumes equal production of $B^0\bar{B}^0$ and $B^+B^-$ pairs.
The systematic error includes the uncertainty in the efficiency,
the number of $B$ mesons (1.4\%), the signal yield (3.7\%)
and the $\psp\to\lp \lm$ branching fraction (2.2\% assuming lepton universality).
This result is combined with the value of the same branching fraction
measured in the $\psp\to J/\psi \pi^+\pi^-$ channel in Ref.~\cite{z4430dalitz},
taking into account the correlations between
the error sources. The final combined result is
\begin{equation*}
\br(\decay) = (5.80\pm0.39)\times10^{-4},
\end{equation*}
where the uncertainty includes statistical and systematic errors.
As we perform a full amplitude analysis, the contributions of the individual
resonances are described more precisely than in Ref.~\cite{z4430dalitz},
and we do not combine the results of the measurements below.

The fit fraction of a resonance $R$ [the $\zm{4430}$ or one of the
$K^*$ resonances]
is defined as
\begin{equation}
f = \frac{\int S_R(\Phi) d\Phi}
{\int S(\Phi) d\Phi},
\end{equation}
where $S_R(\Phi)$ is the signal density function with all contributions other than
the contribution of the $R$ resonance set to 0.
The statistical uncertainties in the fit fractions are determined
from a set of MC pseudoexperiments generated in accordance with the 
fit result in data. We fit each sample and calculate the fit fractions;
the resulting distribution of the fit fractions is fitted to
a Gaussian function,
and the sigma of the Gaussian function is treated as the statistical
uncertainty.
The results are summarized in Table~\ref{tab:ffrac}.

Using the fit fraction of the $\kst(892)$ and the combined $\br(\decay)$,
we calculate the branching fraction of $B^0\to\psp\kst(892)$ decay:
\begin{equation*}
\br(B^0\to\psp\kst(892)) =
(5.55^{+0.22 +0.41}_{-0.23 -0.84})\times10^{-4}.
\end{equation*}
The central value is given for the default model with the $\zm{4430}$
having $J^P = 1^+$.
The systematic error includes contributions from the same sources
as the uncertainty in
the branching fraction of the $\decay$ decay, 
the amplitude model [$(^{+4.8}_{-13.0})\%$] and
the background parametrization [$(^{+0.8}_{-5.5})\%$] dependence
of the $\kst(892)$ fit fraction.
We also determine the fraction
of the $K^*(892)$ mesons that are longitudinally polarized: 
$f_L = (45.5^{+3.1 +1.4}_{-2.9 -4.9})\%$.

The branching fraction product for the $\zm{4430}$ is
\begin{equation*}
\begin{split}
\br(&B^0\to\zm{4430}\kp)\times\br(\zm{4430}\to\psp\pim) = \\
 &(6.0^{+1.7 +2.5}_{-2.0 -1.4})\times10^{-5}, \\
\end{split}
\end{equation*}
where the systematic error due to the amplitude model dependence is
$(^{+41.2}_{-22.4})\%$ and the systematic error due to the
background parametrization dependence is $(^{+3.1}_{-3.5})\%$.


\section{Conclusions}

We have performed an amplitude analysis of $\decay$ decays
in four dimensions.
The preferred assignment of the quantum numbers of the $\zm{4430}$ is $1^+$.
The $J^P=1^+$ hypothesis is favored over the $0^-$, $1^-$, $2^-$ and $2^+$ hypotheses at the levels of $3.4\sigma$, $3.7\sigma$, $4.7\sigma$ and $5.1\sigma$,
respectively. The results for the mass and the width of the $\zm{4430}$ are
\begin{equation*}
\begin{split}
M & = 4485^{+22 +28}_{-22 -11}\ \mevcc, \\
\Gamma & = 200^{+41 +26}_{-46 -35}\ \mev.\\
\end{split}
\end{equation*}
We calculate the branching fractions to be
\begin{equation*}
\begin{split}
& \br(\decay) = (5.80\pm0.39)\times10^{-4}, \\
& \br(B^0\to\psp\kst(892)) =
(5.55^{+0.22 +0.41}_{-0.23 -0.84})\times10^{-4}, \\
\br(&B^0\to\zm{4430}\kp)\times\br(\zm{4430}\to\psp\pim) = \\
&\qquad(6.0^{+1.7 +2.5}_{-2.0 -1.4})\times10^{-5}, \\
\end{split}
\end{equation*}
and the fraction of the longitudinally polarized $K^*(892)$ mesons to be
\begin{equation*}
f_L = (45.5^{+3.1 +1.4}_{-2.9 -4.9})\%.
\end{equation*}
These results supersede previous measurements from
a Dalitz analysis of the same decay channel~\cite{z4430dalitz}.


\section{Acknowledgments}

We thank the KEKB group for the excellent operation of the
accelerator; the KEK cryogenics group for the efficient
operation of the solenoid; and the KEK computer group,
the National Institute of Informatics, and the 
PNNL/EMSL computing group for valuable computing
and SINET4 network support.  We acknowledge support from
the Ministry of Education, Culture, Sports, Science, and
Technology (MEXT) of Japan, the Japan Society for the 
Promotion of Science (JSPS), and the Tau-Lepton Physics 
Research Center of Nagoya University; 
the Australian Research Council and the Australian 
Department of Industry, Innovation, Science and Research;
Austrian Science Fund under Grant No. P 22742-N16;
the National Natural Science Foundation of China under
contract No.~10575109, 10775142, 10875115 and 10825524; 
the Ministry of Education, Youth and Sports of the Czech 
Republic under contract No.~MSM0021620859;
the Carl Zeiss Foundation, the Deutsche Forschungsgemeinschaft
and the VolkswagenStiftung;
the Department of Science and Technology of India; 
the Istituto Nazionale di Fisica Nucleare of Italy; 
The BK21 and WCU program of the Ministry Education Science and
Technology, National Research Foundation of Korea Grant No.\ 
2010-0021174, 2011-0029457, 2012-0008143, 2012R1A1A2008330,
BRL program under NRF Grant No. KRF-2011-0020333,
and GSDC of the Korea Institute of Science and Technology Information;
the Polish Ministry of Science and Higher Education and 
the National Science Center;
the Ministry of Education and Science of the Russian
Federation, the Russian Federal Agency for Atomic Energy and
Russian Foundation for Basic Research grant 12-02-00862-a;
the Slovenian Research Agency;
the Basque Foundation for Science (IKERBASQUE) and the UPV/EHU under 
program UFI 11/55;
the Swiss National Science Foundation; the National Science Council
and the Ministry of Education of Taiwan; and the U.S.\
Department of Energy and the National Science Foundation.
This work is supported by a Grant-in-Aid from MEXT for 
Science Research in a Priority Area (``New Development of 
Flavor Physics''), and from JSPS for Creative Scientific 
Research (``Evolution of Tau-lepton Physics'').


\section{Appendix: derivation of the signal density function}

\subsection{Decay via the $\kst$ resonances}

\begin{figure}[ht]
\includegraphics[width=6cm]{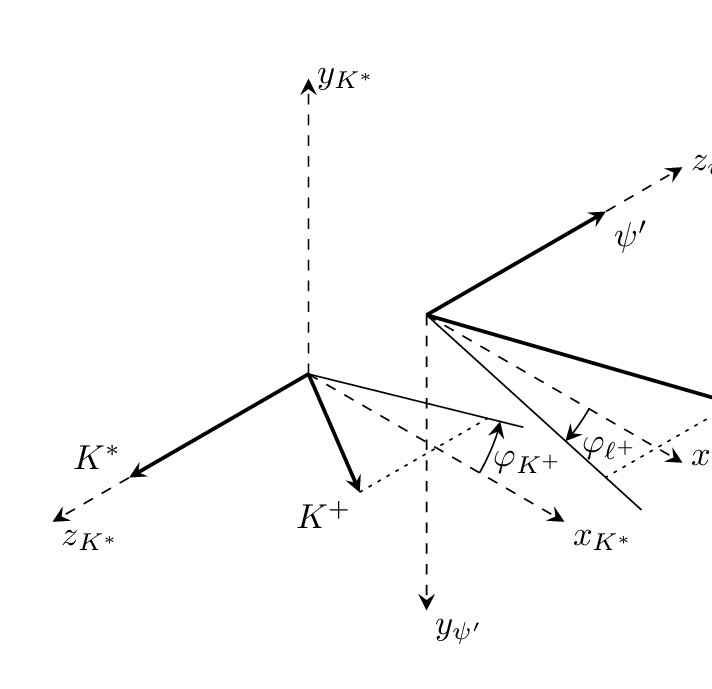}
\caption{Definition of the angle between the decay planes for $\decaykstar$ decay (in the $\B$ rest frame).}
\label{fig:kstplang}
\end{figure}

The definition of the angle between the decay planes of the $\psp$ and $\kst$
is shown in Fig.~\ref{fig:kstplang}.
The coordinate systems
$(x_\kst,y_\kst,z_\kst)$ and $(x_\psp,y_\psp,z_\psp)$
are defined in the $\B$ rest frame; the $x_\kst$ and $x_\psp$ axes
are chosen to be the same. The angle $\phi$ is given by
\begin{equation}
\phi = \phi_\kp + \phi_\lp,
\end{equation}
where $\phi_\kp$ and $\phi_\lp$ are the azimuthal angles of
the $\kp$ and $\lp$, respectively. This angle may be calculated as
\begin{equation}
\begin{split}
\cos \varphi & = \frac{({\vec{a}_\kp}\cdot{\vec{a}_\lp})}{|\vec{a}_\kp||\vec{a}_\lp|}, \\
\sin \varphi & = \frac{([\vec{p}_\psp\times\vec{a}_\kp]\cdot{\vec{a}_\lp})}{|\vec{p}_\psp||\vec{a}_\kp||\vec{a}_\lp|}, \\
\end{split}
\end{equation}
where
\begin{equation}
\begin{split}
\vec{a}_\kp & = \vec{p}_\kp - \frac{({\vec{p}_\kp}\cdot{\vec{p}_\kst})}{|\vec{p}_\kst|^2} \vec{p}_\kst, \\
\vec{a}_\lp & = \vec{p}_\lp - \frac{({\vec{p}_\lp}\cdot{\vec{p}_\psp})}{|\vec{p}_\psp|^2} \vec{p}_\psp, \\
\end{split}
\end{equation}
where $\vec{p}_\kp$, $\vec{p}_\kst$, $\vec{p}_\lp$ and $\vec{p}_\psp$
are the momenta of $\kp$, $\kst$, $\lp$ and $\psp$ in the $\B$ rest frame,
respectively.

\begin{figure}[ht]
\includegraphics[width=6cm]{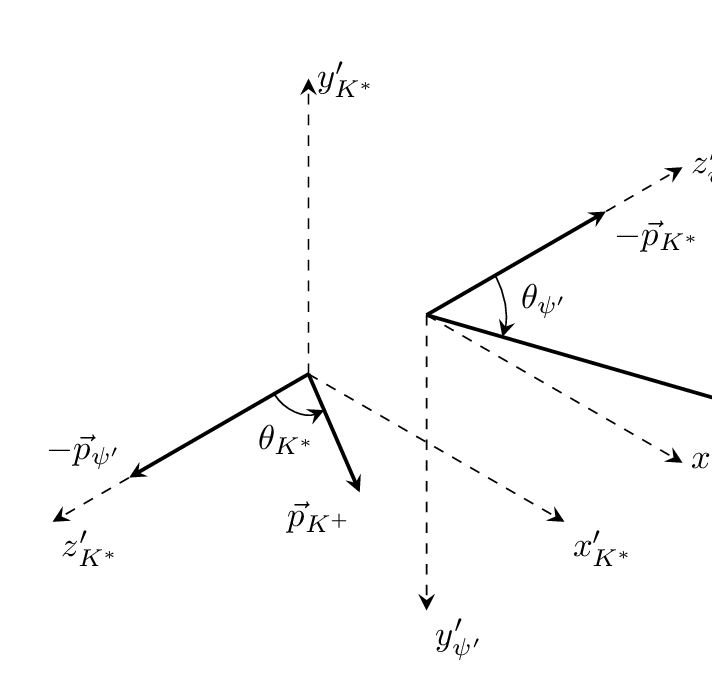}
\caption{Definition of the helicity angles for $\decaykstar$ decay
(in the $\kst$ and $\psp$ rest frames).}
\label{fig:ksthelang}
\end{figure}

The definitions of the helicity angles are shown in Fig.~\ref{fig:ksthelang}.
The coordinate systems $(x'_\kst,y'_\kst,z'_\kst)$ and
$(x'_\psp,y'_\psp,z'_\psp)$ are obtained by the
boosting of the coordinate systems
$(x_\kst,y_\kst,z_\kst)$ and $(x_\psp,y_\psp,z_\psp)$
to the rest frames of the $\kst$ and
$\psp$, respectively. 
The $\kst$ helicity angle is given by
\begin{equation}
\cos \theta_\kst = \frac{-({\vec{p}_\psp}\cdot{\vec{p}_\kp})}{|\vec{p}_\psp||\vec{p}_\kp|},
\end{equation}
where $\vec{p}_\psp$ and $\vec{p}_\kp$ are
the momenta of $\psp$ and $\kp$ in the
$\kst$ rest frame, respectively; the $\psp$ helicity angle is calculated
similarly.

The amplitude of the decay $\decaykstar$ is
\begin{equation}
\small
\begin{split}
A_{\lambda\,\xi}^{\kst}(\Phi)  
& = H^\kst_\lambda
\Dfun{J(\kst)\,*}{\lambda}{0}{\varphi_\kp}{\theta_\kst}{0}
\Dfun{1\,*}{\lambda}{\xi}{\varphi_\lp}{\theta_\psp}{0} \\
& = H^\kst_\lambda
    e^{i\lambda\varphi_\kp} \dfun{J(\kst)}{\lambda}{0}{\theta_\kst}
    e^{i\lambda\varphi_\lp} \dfun{1}{\lambda}{\xi}{\theta_\psp}\\
& = H^\kst_\lambda
    e^{i\lambda\varphi} \dfun{J(\kst)}{\lambda}{0}{\theta_\kst}
      \dfun{1}{\lambda}{\xi}{\theta_\psp},
\end{split}
\label{eq:kstampder}
\end{equation}
where $H^\kst_\lambda$ is the helicity amplitude, $\lambda$ is the helicity of
the $\psp$ and $\xi$ is the helicity of the lepton pair.
Note that the orientation of the coordinate system
$(x''_\psp,y''_\psp,z''_\psp)$ [this is the coordinate system
$(x'_\psp,y'_\psp,z'_\psp)$ rotated by $\varphi_\lp$ around the $z$ axis
and then by $\theta_\psp$ around the $y$ axis]
is fixed by the condition
that the $\kst$ momentum is lying in the plane $(x''_\psp,z''_\psp)$.

\subsection{Decay via the $\zm{4430}$}

\begin{figure}[ht]
\includegraphics[width=6cm]{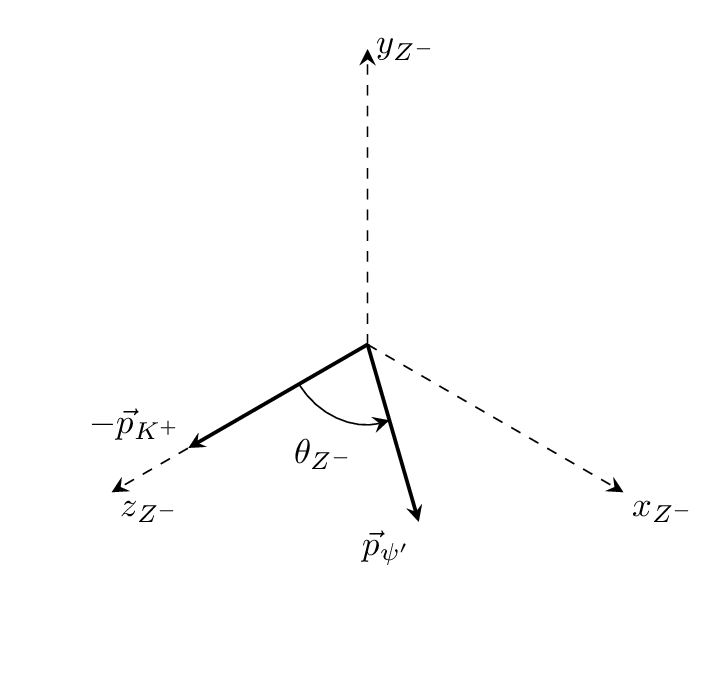}
\caption{Definition of the $\zm{4430}$ helicity angle (in the $\zm{4430}$
rest frame).}
\label{fig:zhelang}
\end{figure}

The definition of the $\z$ helicity angle is shown in Fig.~\ref{fig:zhelang}.
The coordinate system $(x_\z,y_\z,z_\z)$ is defined in the $\zm{4430}$
rest frame and its orientation is chosen so that the $\psp$ momentum is
lying in the plane $(x_\z,y_\z)$.

\begin{figure}[ht]
\includegraphics[width=6cm]{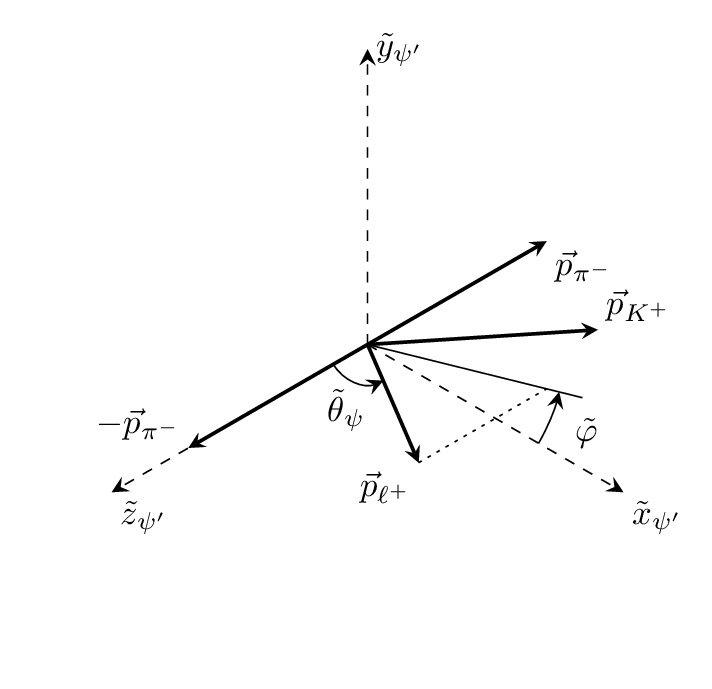}
\caption{Definitions of the $\psp$ helicity angle and
the angle $\tilde{\varphi}$ (the $\psp$ rest frame).}
\label{fig:zpsphelang}
\end{figure}

The definitions of the $\psp$ helicity angle and the angle $\tilde{\varphi}$
are shown in Fig.~\ref{fig:zpsphelang}. The coordinate system
$(\tilde{x}_\psp,\tilde{y}_\psp,\tilde{z}_\psp)$ is defined in the $\psp$
rest frame; the $\kp$ momentum is lying in the plane
$(\tilde{x}_\psp,\tilde{z}_\psp)$, the azimuthal angle being equal to 0.
The azimuthal angle $\tilde{\varphi}$ may be calculated as
\begin{equation}
\begin{split}
\cos \tilde{\varphi} & = \frac{({\vec{a}_\kp}\cdot{\vec{a}_\lp})}{|\vec{a}_\kp||\vec{a}_\lp|}, \\
\sin \tilde{\varphi} & = \frac{-([\vec{p}_\pim\times\vec{a}_\kp]\cdot{\vec{a}_\lp})}{|\vec{p}_\pim||\vec{a}_\kp||\vec{a}_\lp|}, \\
\end{split}
\end{equation}
where
\begin{equation}
\begin{split}
\vec{a}_\kp & = \vec{p}_\kp - \frac{({\vec{p}_\kp}\cdot{\vec{p}_\pim})}{|\vec{p}_\pim|^2} \vec{p}_\pim, \\
\vec{a}_\lp & = \vec{p}_\lp - \frac{({\vec{p}_\lp}\cdot{\vec{p}_\pim})}{|\vec{p}_\pim|^2} \vec{p}_\pim, \\
\end{split}
\end{equation}
where $\vec{p}_\kp$, $\vec{p}_\pim$ and $\vec{p}_\lp$
are the momenta of $\kp$, $\pim$ and $\lp$ in the $\psp$ rest frame,
respectively.

\begin{figure}[ht]
\includegraphics[width=6cm]{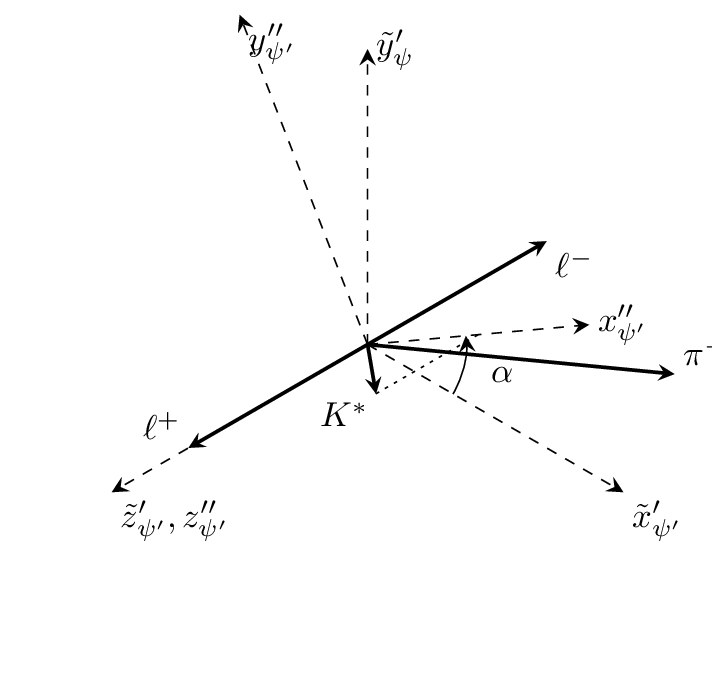}
\caption{Definition of the angle $\alpha$ (in the $\psp$ rest frame).}
\label{fig:psprot}
\end{figure}

The orientation of the coordinate system
$(\tilde{x}'_\psp,\tilde{y}'_\psp,\tilde{z}'_\psp)$
[this is the coordinate system
$(\tilde{x}_\psp,\tilde{y}_\psp,\tilde{z}_\psp)$ rotated by
$\tilde{\varphi}$ around the $z$ axis
and then by $\tilde{\theta}_\psp$ around the $y$ axis]
satisfies the condition that the $\pim$ momentum is lying in the plane
$(\tilde{x}'_\psp,\tilde{z}'_\psp)$; thus, this coordinate system is not the
same as $(x''_\psp,y''_\psp,z''_\psp)$. The coordinate systems in question
are shown in Fig.~\ref{fig:psprot}.
The azimuthal angle $\alpha$ may be calculated as
\begin{equation}
\begin{split}
\cos \alpha & = \frac{({\vec{a}_\pim}\cdot{\vec{a}_\kst})}{|\vec{a}_\pim||\vec{a}_\kst|}, \\
\sin \alpha & = \frac{([\vec{p}_\lp\times\vec{a}_\pim]\cdot{\vec{a}_\kst})}{|\vec{p}_\lp||\vec{a}_\pim||\vec{a}_\kst|}, \\
\end{split}
\end{equation}
where
\begin{equation}
\begin{split}
\vec{a}_\kst & = \vec{p}_\kst - \frac{({\vec{p}_\kst}\cdot{\vec{p}_\lp})}{|\vec{p}_\lp|^2} \vec{p}_\lp, \\
\vec{a}_\pim & = \vec{p}_\pim - \frac{({\vec{p}_\pim}\cdot{\vec{p}_\lp})}{|\vec{p}_\lp|^2} \vec{p}_\lp, \\
\end{split}
\end{equation}
where $\vec{p}_\kst$, $\vec{p}_\pim$ and $\vec{p}_\lp$
are the momenta of $\kst$, $\pim$ and $\lp$ in the $\psp$ rest frame,
respectively. After additional rotation by $\alpha$ around
the $z$ axis, the coordinate system
$(\tilde{x}'_\psp,\tilde{y}'_\psp,\tilde{z}'_\psp)$ becomes the same
as $(x''_\psp,y''_\psp,z''_\psp)$; thus, the final states are the same
for the decays via the $\kst$ and $\zm{4430}$.

The amplitude of the decay $\decayz$ is
\begin{equation}
\begin{split}
A_{\lambda'\,\xi}^{\z}(\Phi)
& = H^\z_{\lambda'}
\Dfun{J(\z)\,*}{0}{\lambda'}{0}{\theta_\z}{0}
\Dfun{1\,*}{\lambda'}{\xi}{\tilde{\varphi}}{\tilde{\theta}_\psp}{\alpha} \\
& = H^\z_{\lambda'}
    \dfun{J(\z)}{0}{\lambda'}{\theta_\z} e^{i\lambda'\tilde{\varphi}}
    \dfun{1}{\lambda'}{\xi}{\tilde{\theta}_\psp} e^{i\xi\alpha}, \\
\end{split}
\label{eq:zampder}
\end{equation}
where $H^\z_{\lambda'}$ is the helicity amplitude and $\lambda'$
is the helicity of the $\psp$.
The amplitudes in Eq.~\eqref{eq:zampder} are related
by parity conservation in the decay $\z \to \psp \pim$:
\begin{equation}
 H_{\lambda'}^{\z} = -P(\z)(-1)^{J(\z)} H^{\z}_{-\lambda'}. \\
\end{equation}
Note that the amplitudes in Eq.~\eqref{eq:kstampder} for $\lambda$ and
$-\lambda$ are not related, because the $\psp$ is produced in the weak decay
$\B \to \psp \kst$.

\subsection{The signal density function}

Combining the amplitudes in Eqs.~\eqref{eq:kstampder} and \eqref{eq:zampder}, one
gets the signal density function for $\decay$ decays:
\begin{equation}
\small
\begin{split}
S(\Phi) = 
\sum_{\xi=1,-1} \left|
\sum_{\kst} \sum_{\lambda=-1,0,1} A_{\lambda\,\xi}^{\kst} +
\sum_{\lambda'=-1,0,1} A_{\lambda'\,\xi}^{\z}
\right|^2.
\end{split}
\label{eq:sigpdfder}
\end{equation}
The lepton pair is produced in the electromagnetic decay $\psp \to \lpair$ via
a virtual photon; thus its helicity $\xi$ may be equal to $1$ or $-1$.

For the charge conjugate decay $\cdecay$,
the particles in the definitions of the angular
variables change to the corresponding antiparticles ($\kp\to\km$,
$\pim\to\pip$, $\lp\to\lm$ and $\lm\to\lp$). If the parity transformation
is applied, then the helicity angles do not change and the azimuthal angles
change sign (because $\cos\tilde{\varphi} \to \cos\tilde{\varphi}$ and
$\sin\tilde{\varphi} \to -\sin\tilde{\varphi}$). Thus,
the signal density for the decay $\cdecay$ is given by
Eq.~\eqref{eq:sigpdfder} with the opposite sign of the azimuthal angles
($\varphi \to -\varphi$, $\tilde{\varphi} \to -\tilde{\varphi}$ and
$\alpha \to -\alpha$).

\end{document}